\documentstyle{camera}

\newcommand{\AmS}{{\protect\the\textfont2
  A\kern-.1667em\lower.5ex\hbox{M}\kern-.125emS}}

\newcommand{\beq}{\begin{equation}}
\newcommand{\eeq}{\end{equation}}
\newcommand{\bea}{\begin{eqnarray}}
\newcommand{\eea}{\end{eqnarray}}

\input epsf

\begin{document}

%
\title{Cen A persistence and Virgo absence versus updated maps\\
  to understand UHECR nature}

%
\author{DANIELE FARGION and DANIELE D'ARMIENTO}

%
\organization{Physics Department and INFN roma1, Univ. Sapienza, Ple. A. Moro 2, 00185, Roma, Italy}

\maketitle

\begin{abstract}
Ultra High Cosmic Rays (UHECR) should be tracing their sources, making a new astronomy. Their events counting are finally growing, by Auger experiment, into cosmic sky. Their map
should follow the mass distribution in a narrow cosmic volume (the GZK cut off region, correlated with Super Galactic Plane (SPG) or Local Group) if they were protons, as most expected. Indeed at first  \cite{Auger-Nov07}
UHECR did seem to follow the GZK cut off and to correlate with SGP, (even if a few of us disagreed \cite{Fargion2008},\cite{Gorbunov09}). Recently \cite{Auger10} the last $69$ UHECR did not longer follow the SGP map, opening the way to very different correlations \cite{Auger10}, and extreme bending connection \cite{Semikoz10} ; we reconfirm here our Lightest Nuclei interpretation \cite{Fargion2008},\cite{Fargion09a},\cite{Fargion09b},\cite{Fargion2009},\cite{Fargion2010}, while showing here the last event over different radio,X,Gamma and tens TeV CR maps. The  Virgo Cluster absence, the persistence of UHECR clustering along Cen A, the first triplet along Vela seem to confirm light nuclei UHECR understanding, implying a very narrow Universe view, even partially of galactic origin. UHECR fragments might follow (at half, fourth of the  energy) the same UHECR map with a tail or a crown clustering around  main UHECR group. Also secondary gamma and UHE neutrino might trace partially those maps. Tau neutrinos at EeV or PeVs   may play a role in correlating UHECR map and disentangling nucleon from  nuclei nature, possibly in Auger Fluorescence Telescopes, by night horizontal up-going tau airshowers.

\end{abstract}
\vspace{1.0cm}

\section{Introduction}
UHECR astronomy is a hope becoming a reality, with some confusion because of the magnetic field smearing of the arrival directions. Moreover
  while flying UHECR are making photo-pion (if nucleon) or gamma and neutrinos by photo-dissociation (if light nuclei). Making UHECR nucleon local and sharp (GZK cut off, tens Mpc) or very local and smeared (a few Mpc) for lightest nuclei. Therefore UHECR astronomy is surrounded by a parasite astronomy made by gamma and neutrinos as well as by their possible small radio tails as well as UHECR fragments. Indeed UHECR formed by  lightest nuclei may explain clustering of events around CenA and a puzzling UHECR absence  around Virgo. Their fragments $He + \gamma \rightarrow D+D, He + \gamma \rightarrow He^{3}+n, He + \gamma \rightarrow T +p $  may trace on the same UHECR maps by a secondary tail or a crown clustering at half or fourth the UHECR primary  energy. Neutrinos and gamma are tracing  (both for nucleon or nuclei) their UHECR trajectory, respectively at EeVs or PeV energy. Gamma rays are partially absorbed by microwave and infrared background making only a very local limited astronomy. Among neutrinos $\nu$,  muons ones $\nu_{\mu}$, the most penetrating and easy to detect, are deeply polluted by atmospheric  component (smeared and isotropic like their parent CR ). As shown by last TeV muon neutrino maps  probed by very smooth ICECUBE neutrinos map. Tau neutrinos,  the last neutral lepton discovered, absent in neutrino oscillation at TeVs-PeVs-EeVs atmospheric windows,  may arise as the first clean signal in UHECR-neutrino associated astronomy. Their tau birth in ice  may shine as a double bangs (disentangled above PeV) anisotropy. In addition UHE tau, born tangent to the  Earth or mountain, while escaping  in air  may lead, by decay in flight, to  loud, amplified  well detectable tau-airshower at horizons. Tau astronomy versus UHECR are going to reveal most violent  sky as the most deepest probe. First hint of  Vela, the brightest and nearest gamma source, a first galactic source  is rising as a UHECR triplet nearby. Cen A (the most active and nearby AGN) is apparently shining  UHECR source whose clustering (almost a quarter of the event) along a narrow solid angle around (whose opening angular size is  $\simeq 17^{o}$) seem  firm and it is  favoring lightest nuclei. Remaining  events are possibly more  smeared being more bent and heavier nuclei  of  galactic and-or  extragalactic origin.

The rise of nucleon UHECR above GZK astronomy made by protons (AUGER November 2007) is puzzled by three main mysteries: an unexpected nearby Virgo UHECR suppression (or absence), a rich crowded clustering frozen vertically along Cen A, a composition suggesting nuclei (not much directional) and not nucleons. The UHECR map, initially consistent with  GZK volumes, to day seem to be not much correlated with expected Super Galactic Plane. Moreover slant depth data of UHECR from AUGER airshower shape do not favor the proton but points to  a nuclei. To make even more confusion (or fun) HIRES, on the contrary, seem to favor, but with less statistical weight, UHECR mostly nucleons. We tried  (at least partially) to solve the contradictions assuming UHECR as light nuclei ( $He^4$, Li, Be)) spread by planar galactic fields, randomly at vertical axis. The $He^4$ fragility and its mass and its charge explains naturally the Virgo absence (due to $He^4$ opacity above few Mpc) and the observed wide Cen A spread clustering  (a quarter of the whole sample within $17^{o}$). However more events and rarest doublets and  clustering are waiting for an answer. Here we foresee hint of a new UHECR component, due to a first timid triplet toward Vela, the nearest and the brightest gamma source (at few or tens GeV) as well as the possible first high energy source (cosmic ray at tens TeV). Indeed early anisotropy of tens TeVs Cosmic rays found in ICECUBE muons might confirm this possibility.

Let us remind that last century have seen the birth of a puzzling cosmic ray whose nature and origination has been and it is still growing in an apparent never-ending chain of  puzzle.
The Cosmic Black Body Radiation had imposed since $1966$ a cut, GZK cut-off \cite{Greisen:1966jv}, of highest energy cosmic ray propagation, implying a very limited cosmic Volume (ten or few tens Mpc) for highest UHECR  nucleon events. Because of the UHECR rigidity one had finally to expect to track easily UHECR directionality back toward the sources into a new Cosmic Rays Astronomy.
Indeed in last two decades, namely since 1991-1995 the rise of an \emph{apparent} UHECR at $3 {10^{20}}$ eV, by Fly's Eye,  has opened the wondering of its origination: no nearby (within GZK cut off) source have been correlated. Incidentally it should be noted that even after two decades  and after an increase of aperture observation  by nearly two order of magnitude (area-time by AGASA-HIRES-AUGER) no larger or equal event as $3 {10^{20}}$ eV has been rediscovered. Making wondering the nature of that exceptional starting UHECR event. To face the uncorrelated UHECR at $3 {10^{20}}$ eV, and later on event by AGASA, the earliest evidences (by SuperKamiokande) that neutrino have a non zero mass had opened  \cite{Fargion1997} the possibility of  an UHECR-Neutrino connection:  UHE ZeV neutrino could be the transparent  courier of a far AGN (beyond GZK radius) that may hit and scatter on  a local relic anti-neutrino dark halo, spread  at a few Mpc around our galaxy. Its resonant Z boson (or WW channel) production  \cite{Fargion1997} is source of a secondary nucleon later on observable at Earth as a UHECR.  The later search by AGASA seemed to confirm the Fly's Eye by events above $ {10^{20}}$eV  and the puzzling absence of nearby expected GZK anisotropy or correlation.  On $1999-2000$ we were all convinced on the UHECR GZK cut absence. In different occasion HIRES data offered a possible  UHECR connections with far BL Lac \cite{Gorbunov}, giving argument to such Z-resonant (Z-burst) model. However more recent  records by a larger Hires area  (2001-2005)  have been claiming  evidences of GZK suppression in UHECR spectra. The same result seemed confirmed by last AUGER data in last few years \cite{Auger-Nov07}. But in addition AUGER have shown an anisotropic clustering, seeming along the Super Galactic Plane, a place well consistent with GZK expectation \cite{Auger-Nov07}. Because of it the UHE neutrino scattering model \cite{Fargion1997},\cite{Weiler1997},\cite{Yoshida1998}) became obsolete.
Recent  updated maps didn't solve in our opinion the Auger puzzle : indeed several catalog might be correlated with UHECR \cite{Auger10} , but not all of them, (in particular gamma ones whose redshift are unknown)  are located within a GZK volume \cite{Auger10}: moreover a very speculative and alternative explanation (by one of the Auger group author \cite{Semikoz10}) is calling for a cooperative bending, by extragalactic and galactic magnetic field, of the UHECR from Virgo cluster overlapping on the same Cen A area, making there a group of event crowding. We do not share this tuned explanation that do not explain the narrow angle events overlapping to Cen A .
 Indeed here we review the most recent  UHECR maps \cite{Auger10} over many different Universe \emph{colors} or (wave-length)  and we comment some  feature, noting some possible minor galactic component \cite{Fargion09b}, \cite{Fargion2010}. Moreover we remind possible Z-Showering model solution if UHECR are correlated to AGN,BLac or Quasars at large redshift. The consequences of the UHECR composition and source reflects into UHE (GZK \cite{Greisen:1966jv} or cosmo-genic) neutrinos. The proton UHECR provide EeV neutrinos (muons and electron) whose flavor oscillation lead to tau neutrinos to be soon detectable \cite{FarTau} \cite{Auger08} by upward tau air-showers;  the UHECR lightest nuclei model provide only lower energy, tens PeV, neutrinos detectable in a very peculiar way by AUGER fluorescence telescopes or in ARGO array by horizontal $\tau$ air-showers  , or by Icecube $km^3$ neutrino fluorescence telescopes \cite{FarTau},\cite{Fargion2009},\cite{Fargion09a}  \cite{Fargion09b} either by double bang\cite{Learned}, or long muon at few PeV energy. ZeV UHE neutrinos in Z-Shower model are possible source of horizontal Tau air-showers of maximal size and energy \cite{FarTau},\cite{Fargion2010}.
\section{The Lorentz UHECR bending and spread}
Cosmic Rays are blurred by magnetic fields. Also UHECR suffer of a Lorentz force deviation. This smearing maybe source of UHECR features. Mostly along Cen A.
 There are two main spectroscopy of UHECR along galactic plane:
 A late nearby (almost local) bending by a nearest coherent galactic arm field, and a random one along the whole plane inside different arms.
The coherent Lorentz angle bending $\delta_{Coh} $ of a proton UHECR (above GZK \cite{Greisen:1966jv}) within a galactic magnetic field  in a final nearby coherent length  of $l_c = 1\cdot kpc$ is $ \delta_{Coh-p} \simeq{2.3^\circ}\cdot \frac{Z}{Z_{H}} \cdot (\frac{6\cdot10^{19}eV}{E_{CR}})(\frac{B}{3\cdot \mu G}){\frac{l_c}{kpc}}$.
The corresponding coherent  bending of an Helium UHECR at same energy, within a galactic magnetic field
  in a wider nearby coherent length  of $l_c = 2\cdot  kpc$ is
\begin{equation}
\delta_{Coh-He} \simeq
{9.2^\circ}\cdot \frac{Z}{Z_{He}} \cdot (\frac{6\cdot10^{19}eV}{E_{CR}})(\frac{B}{3\cdot \mu G}){\frac{l_c}{2 kpc}}
\end{equation}


This bending angle is compatible with observed multiplet along $Cen_A$ and also the possible clustering along Vela, at much nearer distances; indeed in latter case it is possible for a larger magnetic field along its direction (20 $\mu G$) and-or for a rare iron composition $\delta_{Coh-Fe-Vela} \simeq {17.4^\circ}\cdot \frac{Z}{Z_{Fe}} \cdot (\frac{6\cdot10^{19}eV}{E_{CR}})(\frac{B}{3\cdot \mu G}){\frac{l_c}{290 pc}}$. Such iron UHECR are mostly bounded inside a Galaxy, as well as in Virgo, explaining partially  its extragalactic absence. In lightest nuclei model the heavier of lightest nuclei that may be bounded from Virgo, Be, is bent by $
\delta_{Coh-Be} \simeq {18.4^\circ}\cdot \frac{Z}{Z_{Be}} \cdot (\frac{6\cdot10^{19}eV}{E_{CR}})(\frac{B}{3\cdot \mu G}){\frac{l_c}{2 kpc}}
$.  The incoherent random angle bending, $\delta_{rm} $, while crossing along the whole Galactic disk $ L\simeq{20 kpc}$  in different spiral arms  and within a characteristic  coherent length  $ l_c \simeq{2 kpc}$ for He nuclei is $\delta_{rm-He} \simeq{16^\circ}\cdot \frac{Z}{Z_{He^2}} \cdot (\frac{6\cdot10^{19}eV}{E_{CR}})(\frac{B}{3\cdot \mu G})\sqrt{\frac{L}{20 kpc}}
\sqrt{\frac{l_c}{2 kpc}}$ The heavier  (but still lightest nuclei) bounded from Virgo are Li and Be:
$\delta_{rm-Li} \simeq {24^\circ}\cdot \frac{Z}{Z_{Li^3}} \cdot (\frac{6\cdot10^{19}eV}{E_{CR}})(\frac{B}{3\cdot \mu G})\sqrt{\frac{L}{20 kpc}}
\sqrt{\frac{l_c}{2 kpc}} $, $\delta_{rm-Be} \simeq{32^\circ}\cdot \frac{Z}{Z_{Be^4}} \cdot (\frac{6\cdot10^{19}eV}{E_{CR}})(\frac{B}{3\cdot \mu G})\sqrt{\frac{L}{20 kpc}}
\sqrt{\frac{l_c}{2 kpc}}$.  It should be noted that the present anisotropy above GZK \cite{Greisen:1966jv} energy $5.5 \cdot 10^{19} eV$ might leave a tail of signals: indeed the photo disruption of He into deuterium, Tritium, $He^3$ and protons (and unstable neutrons), might rise as clustered events at half or a fourth of the energy.  It is important to look for correlated tails of events, possibly in  strings at low $\simeq 1.5-3 \cdot 10^{19} eV$ along the $Cen_A$ train of events. \emph{It should be noticed that Deuterium fragments are half energy and mass of Helium: Therefore D and He spot are bent at same way and overlap into circle clusters.}  Deuterium are even more bounded in a local Universe because their fragility. In conclusion He like UHECR  maybe bent by a characteristic as large as  $\delta_{rm-He}  \simeq 16^\circ$. Well within the observed CenA UHECR clustering spread.


\section{UHECR Maps: a brief tour in multi-wave sky}
\begin{figure}[h!]  
\begin{center}
\epsfysize=4cm \hspace{5.0cm} \epsfbox{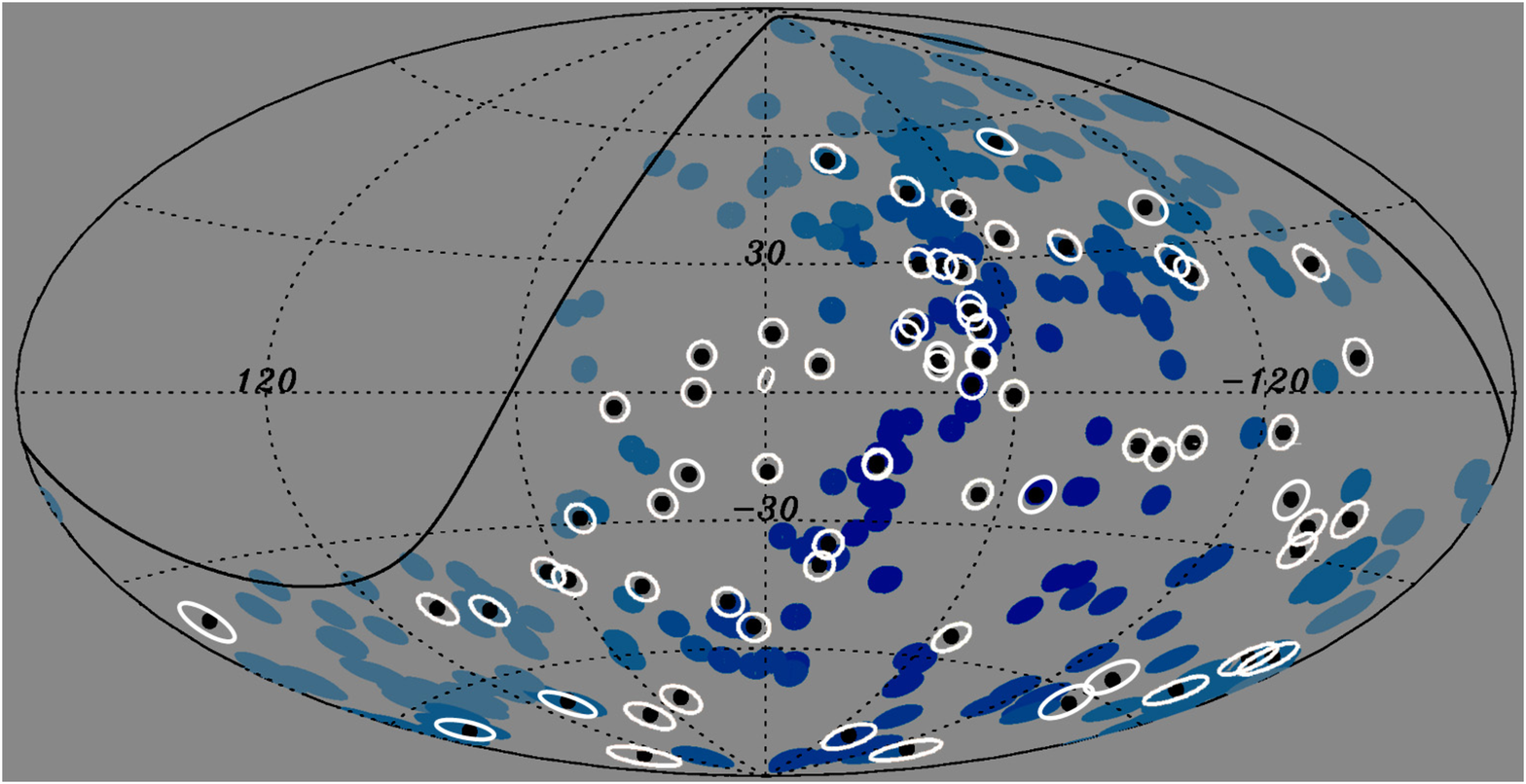}
\epsfysize=5.5cm \hspace{5.0cm} \epsfbox{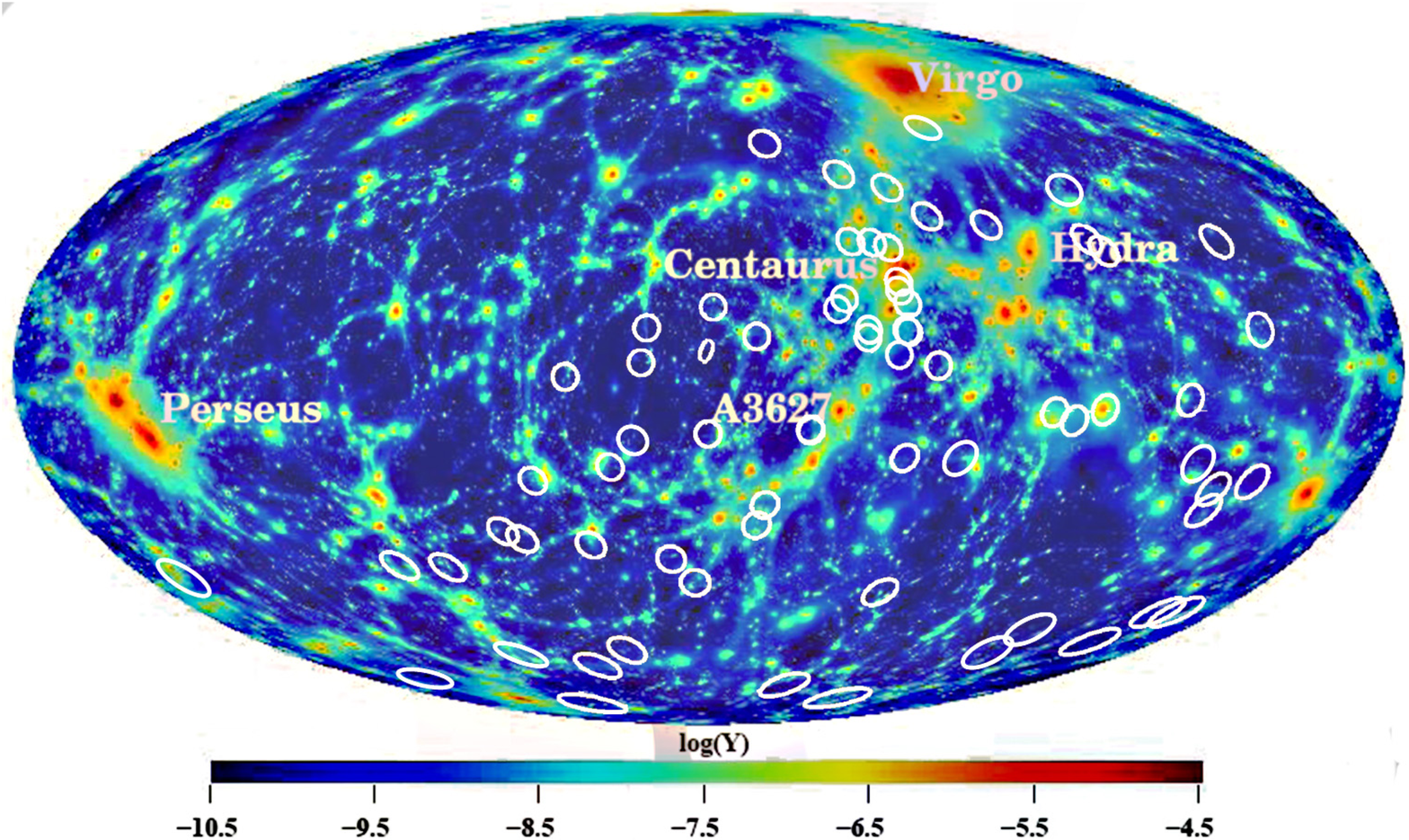}
\epsfysize=4.25cm \hspace{5.0cm} \epsfbox{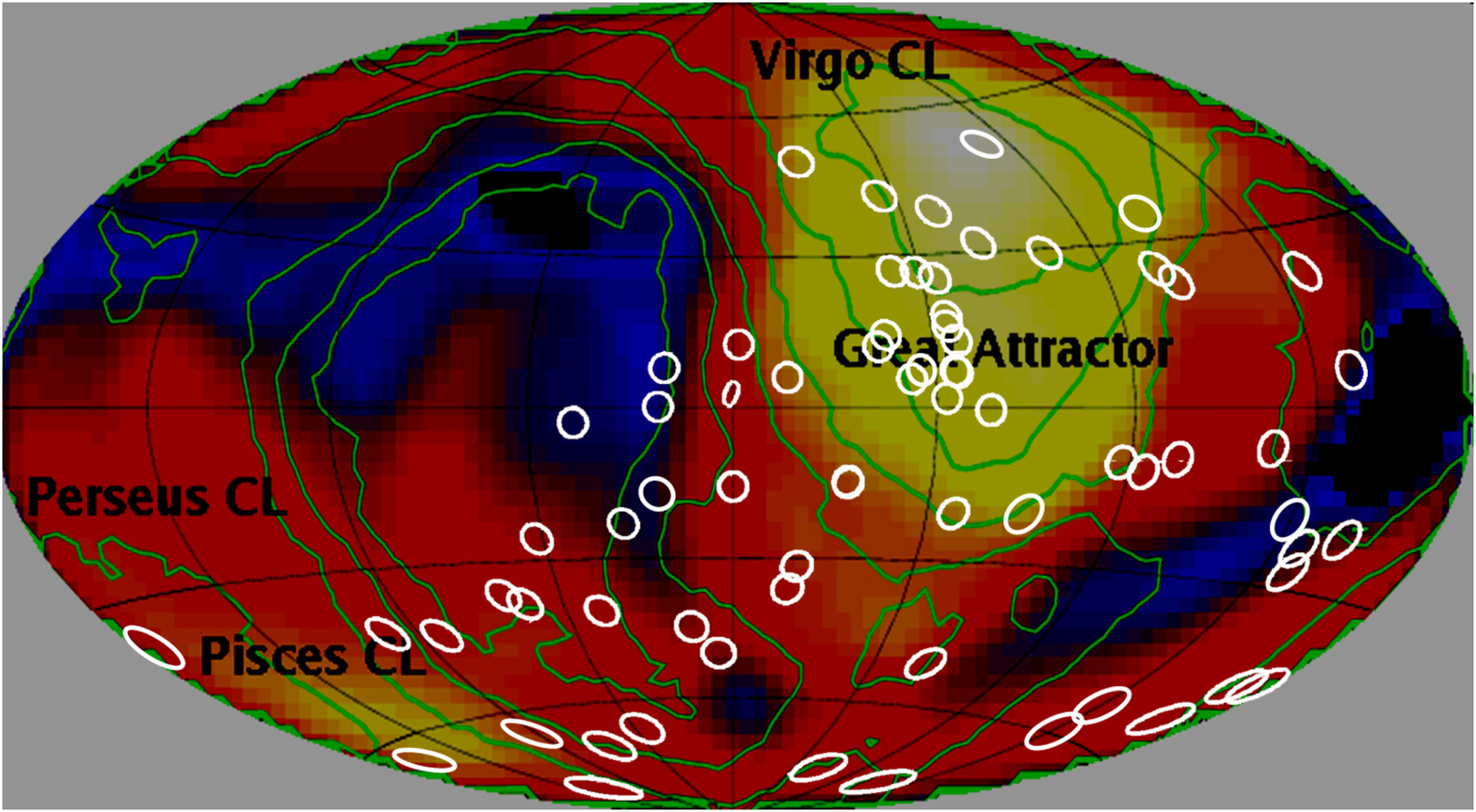}
\caption{ An overlap of last published 69 UHECR maps over different presentation maps. On the first figure: The probability to \emph{not} observe any UHECR around Virgo area center within (one  $3.5^{o}$,two $7^{o}$, or three $10.5 ^{o}$ angle radius) far from each AGN  is respectively $37.15\%, 1.9\%, 0.12\%$, respect to the AGN sources considered by \cite{Auger-Nov07},\cite{Auger10} Auger articles; therefore from these unlike statistics the puzzling paucity of (expected) Virgo sources.  The ring radius is nearly $3.5^o$. No apparent correlation between lights (and common constellation maps)  and UHECR records maybe found. The second map shows a infrared IRAS nearby Universe and the UHECR 69 AUGER events \cite{Auger10}. It is well apparent the bright cluster of Virgo remarked by the absence of any UHECR. The UHECR GZK proton cut-suppression reduces the Virgo UHECR flux by a negligible  factor $\simeq 20\%$, leading to an expected  factor ($\simeq 80\%$ of the original signal)not yet observed. We remind that AUGER Virgo sky is not the best to reveal by AUGER see \protect\cite{Auger-Nov07}. However  a simple statistical Poisson test (based on a sample of $69$ UHECR events counts  at same equiprobable area) imply a small probability  $\simeq 1.9\%$ to not find any signal in Virgo area or even less $(1.2\cdot 10^{-1})\%$ allowing $13.5^{o}$ angular dispersion around each AGN (potential sources); this count is based on the AGN presence in the Virgo direction over other AGN in equiprobable AUGER areas. From this paradox our He-UHECR model explains at best the Virgo absence by severe lightest nuclei opacity, while coexisting with a nearer Cen A UHECR clustering \cite{Fargion2008} \cite{Fargion2009} \cite{Fargion2010}. The Fornax cluster presence might be related to nearer galaxy signals.
The last INTEGRAL X ray map shows the dramatic absence of Virgo and local GZK volume sources (as the Great Wall structure).}
\label{fig1}
\end{center}
\end{figure}

We offer in next figures a map view of the last AUGER UHECR events over the different sky wave-band astronomy. At each step the caption explains and offer the arguments for mysteries and surprises. We begin on figure \ref{fig1} to show our mask calibration based on two AUGER different recent presentation \cite{Auger10}  and the same event are shown on different wave-length sky map in common galactic coordinate. These maps are IRAS ones or Integral X ray one. They show dramatically the Virgo Cluster absence. In the next figure \ref{fig2} we discuss the two micron view and we make a few consideration on a probable galactic (secondary) UHECR component (to be confirmed in a near future). Therefore we overlay these UHECR ring events onto a recent Local Universe Map showing its viability and success to correlate with most events ( figure \ref{fig3}). The Lightest nuclei model naturally explain the need of such a nearby Universe for UHECR.
However the peculiar spread of the events around Cen A makes us believe that vertical spread is not only related to galaxies distribution but also on Lorentz bending. The next  ( figure \ref{fig2}) describe the UHECR over the $408$ Mhz cosmic background, showing the peak role of the Cen A emission and Vela emission with UHECR clustering. Also the Ghz polarized map explain the bending around CenA. IR maps and Planck map as well as H$\alpha$ and dust radio map are used to look for correlations \ref{fig2}.  The following  two maps  ( figure \ref{fig3}, are dealing with infrared " Mass galaxy counting at  $2 \mu$ wavelength. At
\ref{fig4}  we show the oldest  EGRET maps with unidentified and identified sources, labels and some  correlations with UHECR  ( figure \ref{fig4}). The next figure ( figure \ref{fig5}) shows the old Comptel Map correlating (strongly) with UHECR map. The galactic nature of such MeV -UHECR connection is remarkable. Also the  20 TeV Icecube map correlation with UHECR suggest a Vela role in Gamma, Radio, UHECR and tens TeV CR. The nest four figures discuss first the composition (see figure \ref{fig6}) and the consequent GZK volumes for possible UHECR nature ( see Figure \ref{fig7}): nucleons, Light Nuclei, Iron. The surviving fraction of proton with distances as well as the interaction distance for nucleon and lightest nuclei are shown in last two figures of the group (see Figure \ref{fig7}). The final( see Figure \ref{fig8}) overlap the very hard Fermi and TeV sky with UHECR events: a clear Cen A role is blowing, but  also few far AGN or BL Lac are correlating, whoever well above GZK cut-off. In particular in last figure ( see Figure \ref{fig8}) it is shown the last (second December 2009) dramatic bright gamma shining of the AGN  $3C454.3$ whose last flare at cosmic  edge ($2 Gpc$) and its eventual correlation might force us to a Z-Shower solution \cite{Fargion1997}.

\begin{figure}[!ht]  
\begin{center}
\epsfysize=3.5cm \hspace{0.5cm} \epsfbox{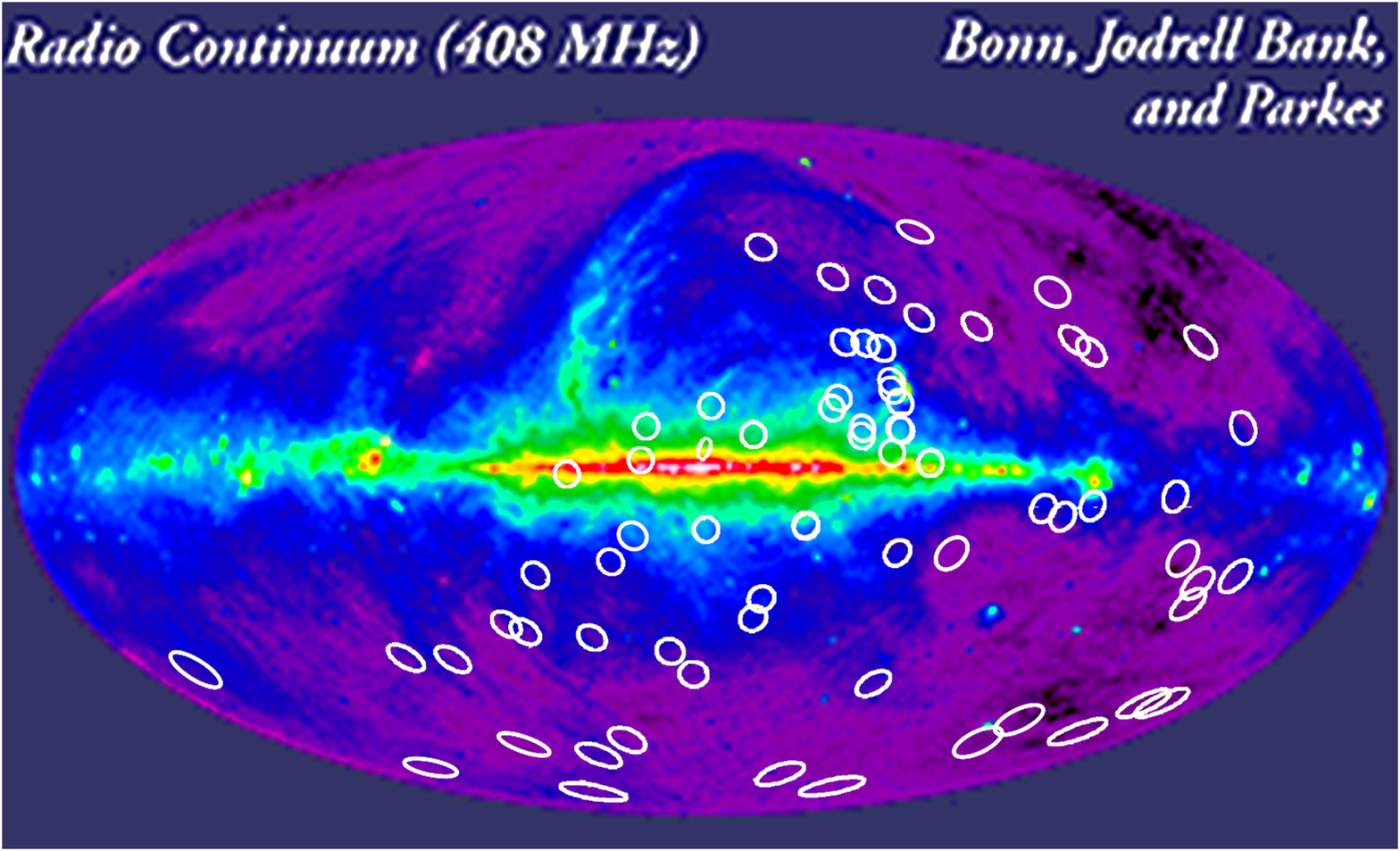}
\epsfysize=3.7cm \hspace{0.5cm} \epsfbox{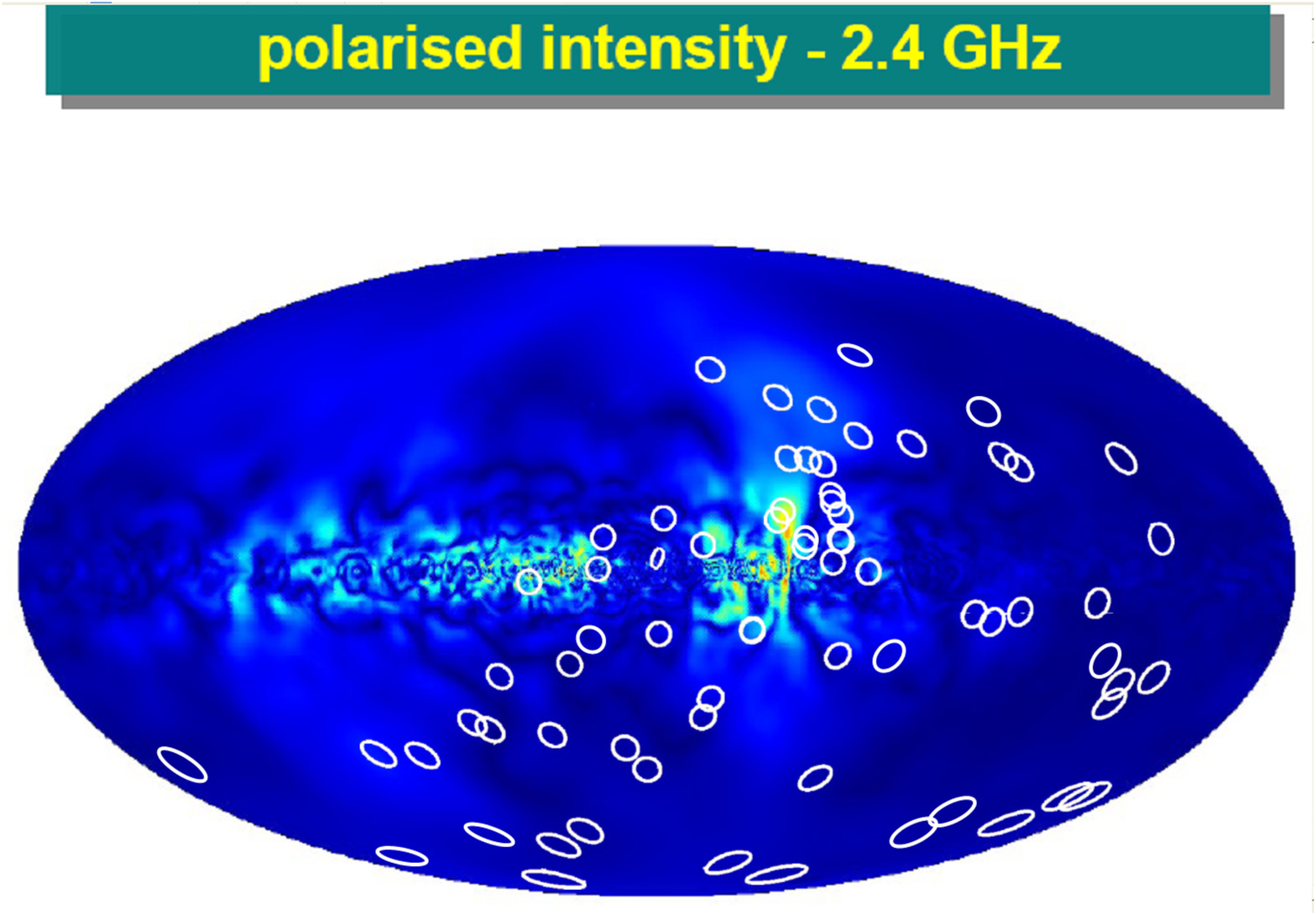}
\epsfysize=3.6cm \hspace{0.5cm} \epsfbox{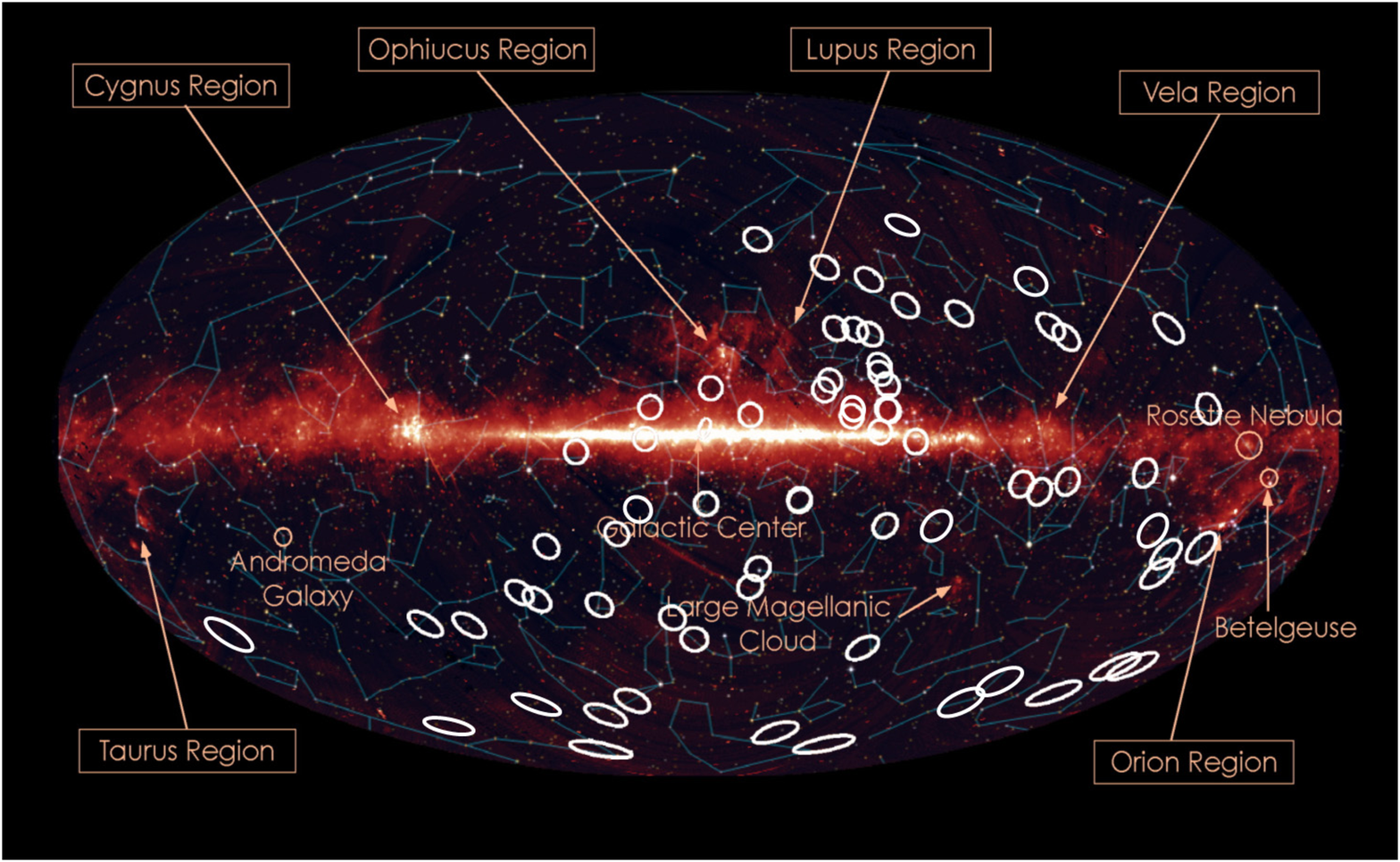}
\epsfysize=3.8cm \hspace{0.5cm} \epsfbox{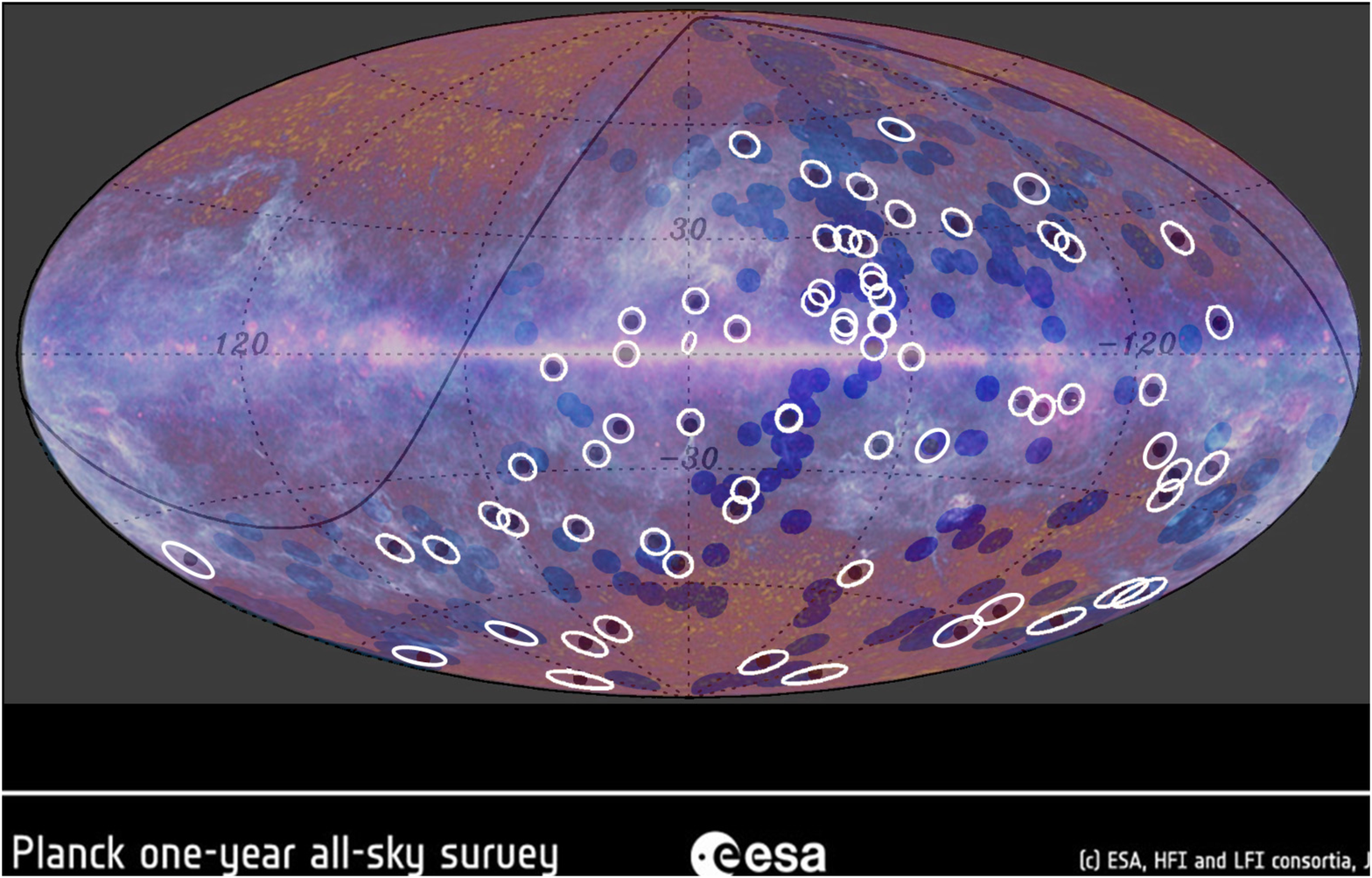}
\epsfysize= 3.7cm \hspace{0.4cm} \epsfbox{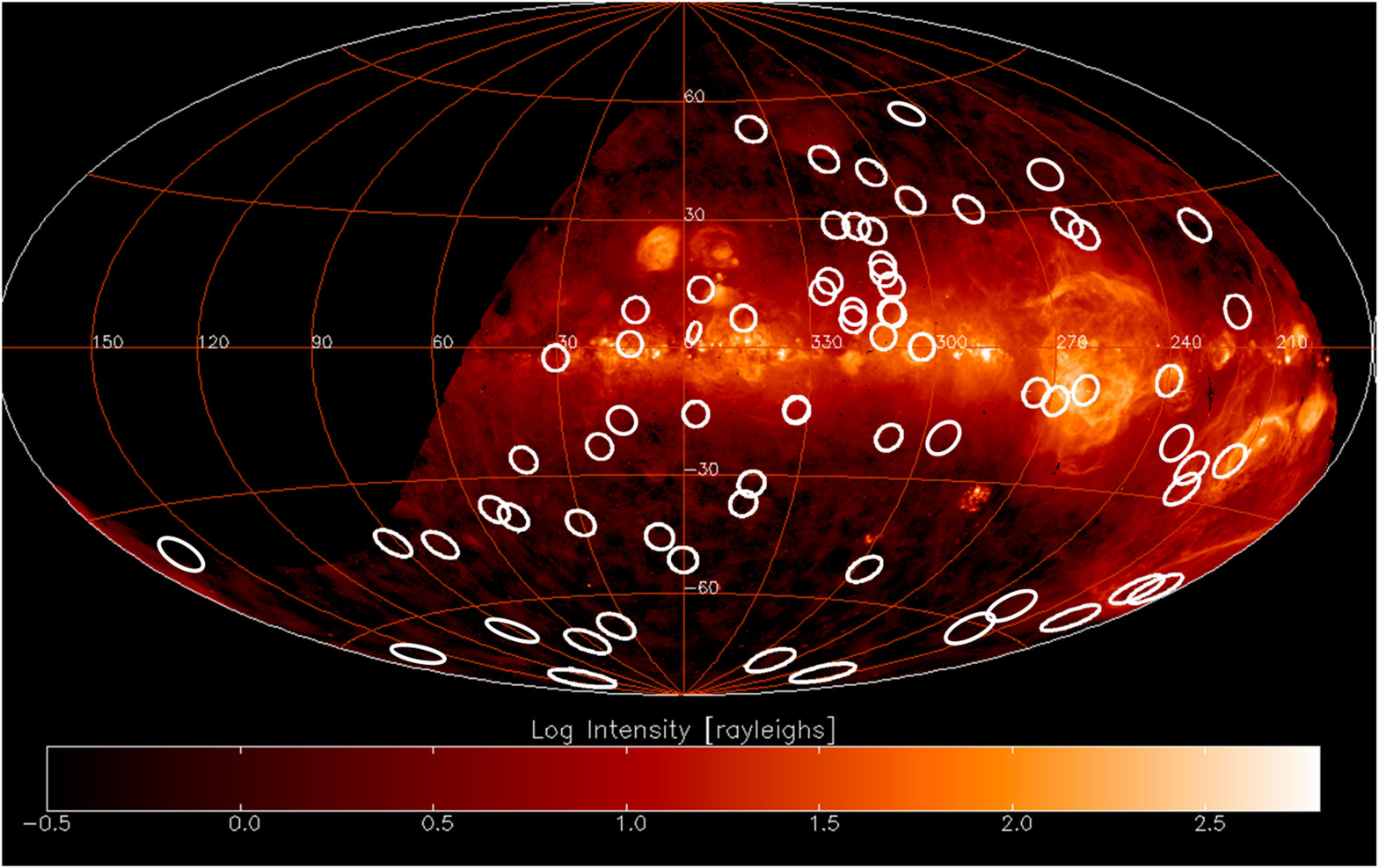}
\epsfysize= 3.6cm \hspace{0.4cm} \epsfbox{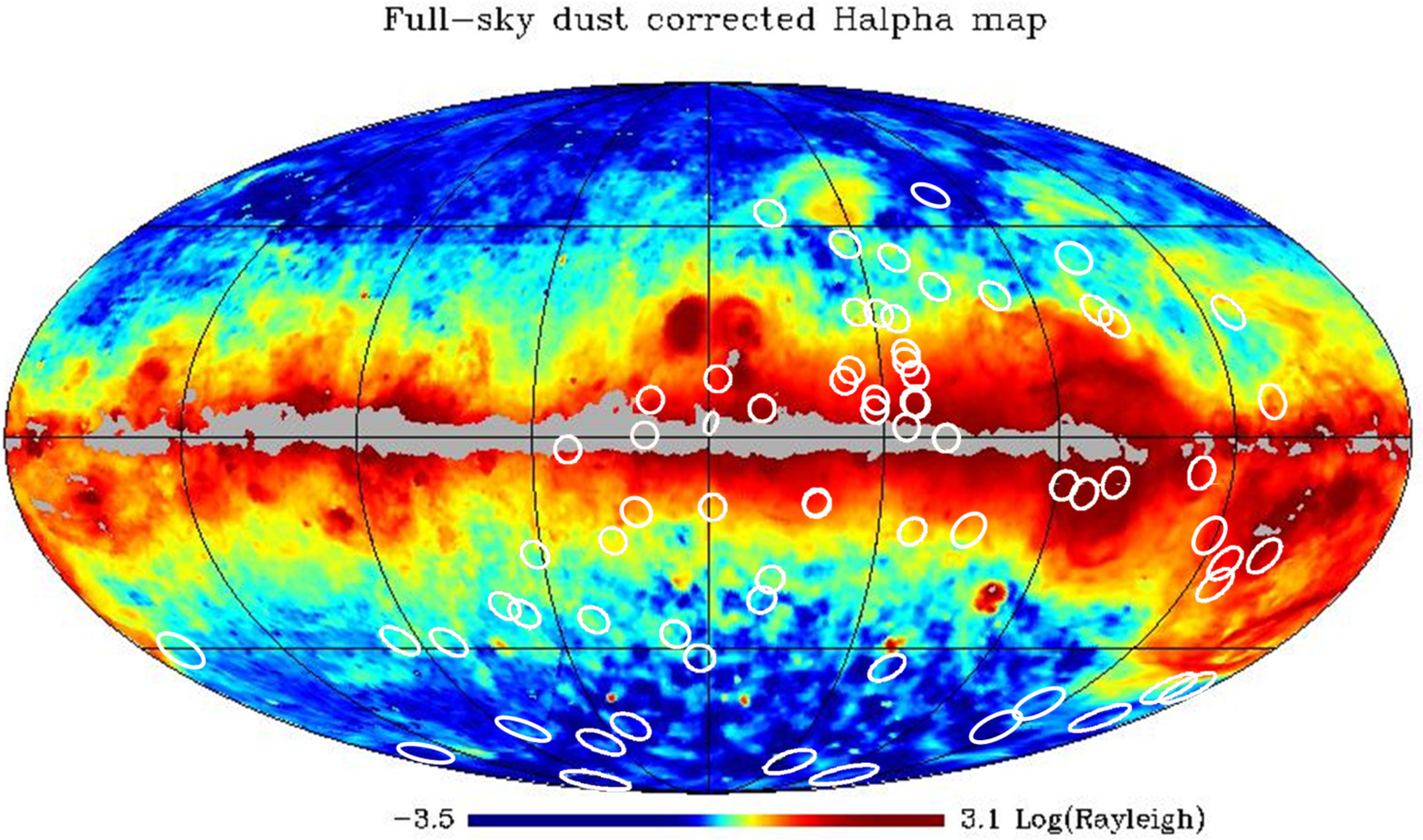}
\end{center}
\caption{The $408$ MHz radio sky whose unique nearest extragalactic and brightest radio source, CenA, is overlapping with the unique evident clustering at galactic plane edge. This clustering is the main significant message we learn from AUGER $69$ \cite{Auger10} UHECR map, both in 2007 and 2010 records. The peak radio activity of Cen A (and partially in gamma band) and its UHECR clustering hint a similar role for nearest Vela PSR, whose radio and gamma activity is, in our galactic plane, the brightest and the extreme one. Indeed we notice and underline a possible correlation of a triplet along Vela whose composition might be a iron or a He strongly bent by arm galactic magnetic field see \protect\cite{Fargion09a},\protect\cite{Fargion09b} The next radio Map at 2.4 Ghz shows an interesting polarization tail along Cen A UHECR clustering. The JAXA satellite AKARI Infrared map is the background of the third figures with marginal correlation with UHECR; the fourth Planck map in microwave band shows an interesting feature: most of the UHECR events seat on the white area associated with the synchrotron radiation around galactic plane and partial halo suggesting, at least, a partial galactic origin of some UHECR. The  fifth and sixth map respectively show the microwave Hydrogen lines and the dust IR radiation, whose correlation with UHECR are not guaranteed. }
\label{fig2}

\end{figure}

\begin{figure}[!ht]  
\begin{center}
\epsfysize=5cm \hspace{0.4cm} \epsfbox{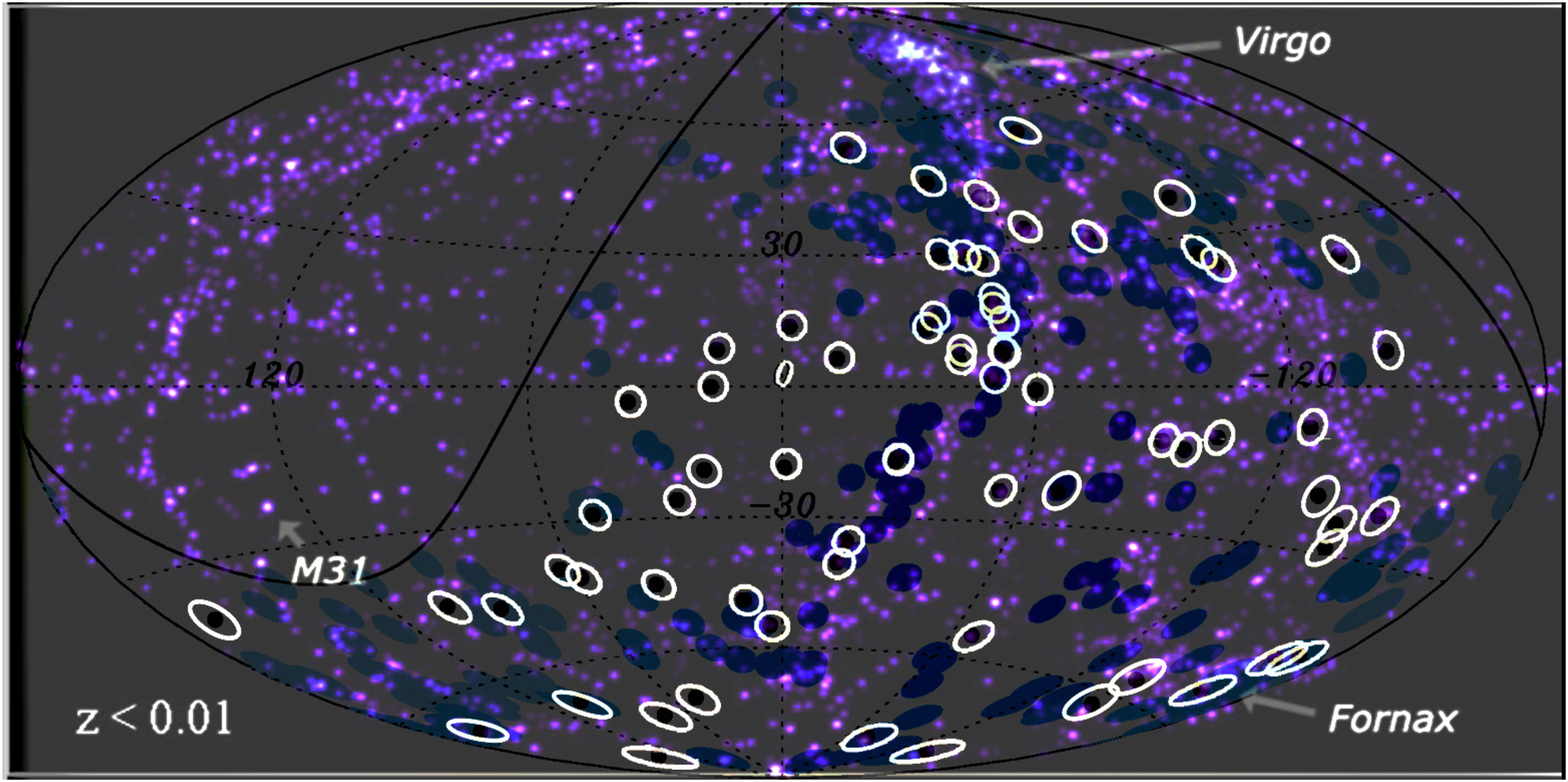}
\epsfysize=5cm \hspace{0.4cm} \epsfbox{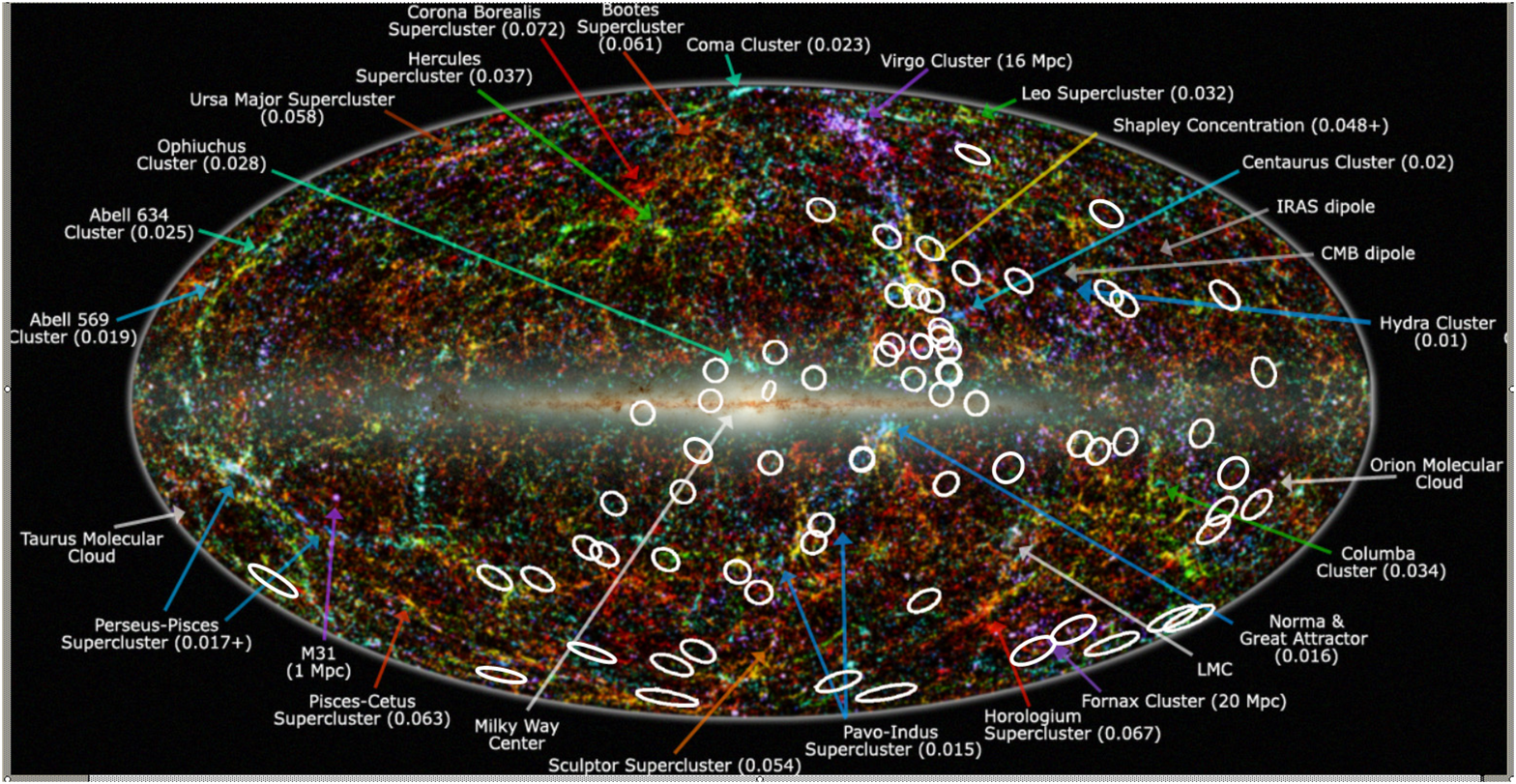}
\vspace{-0.3cm}
\end{center}
\caption{ The analogous IR map by 2MASS LSS chart overlapping on 69 Auger \cite{Auger10} show the absence of Virgo and the persistence of CenA (nearest AGN at 4 Mpc) leading, as shown in the text to favor He-like UHECR nuclei. The second figure show that there are no UHECR events clustering along any GZK structure at redshift $z < 0.01-0.02$, (super galactic plane, Great Wall, Norma Cluster).
The eventual clustering toward Centaurus cluster and Shapley concentration at large distances (a hundred Mpc) is suppressed by GZK cutoff.
}
\label{fig3}
\end{figure}


\begin{figure}[h!]  
\begin{center}
\epsfysize=5cm \hspace{4.0cm} \epsfbox{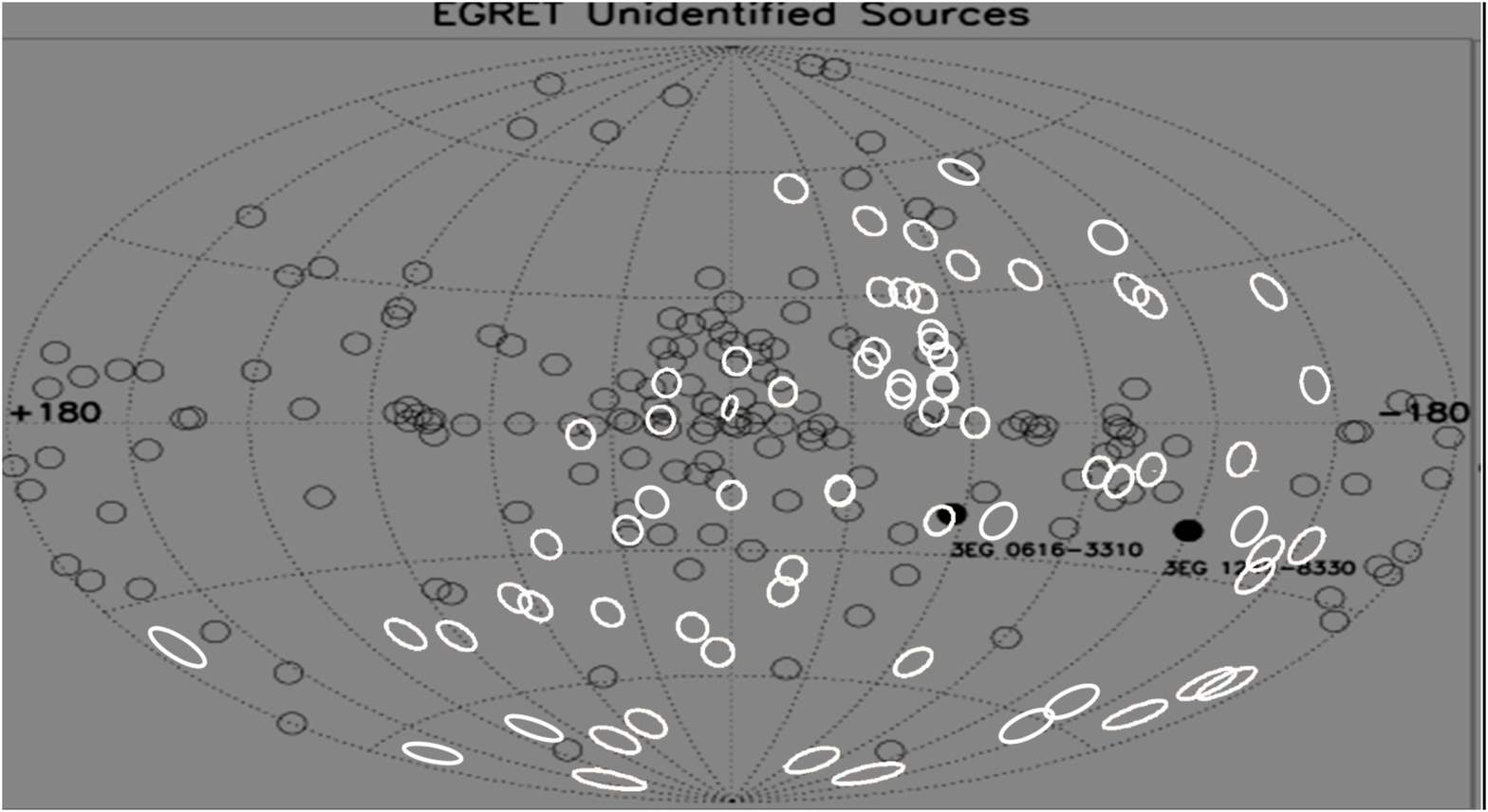}
\epsfysize=5cm \hspace{4.0cm} \epsfbox{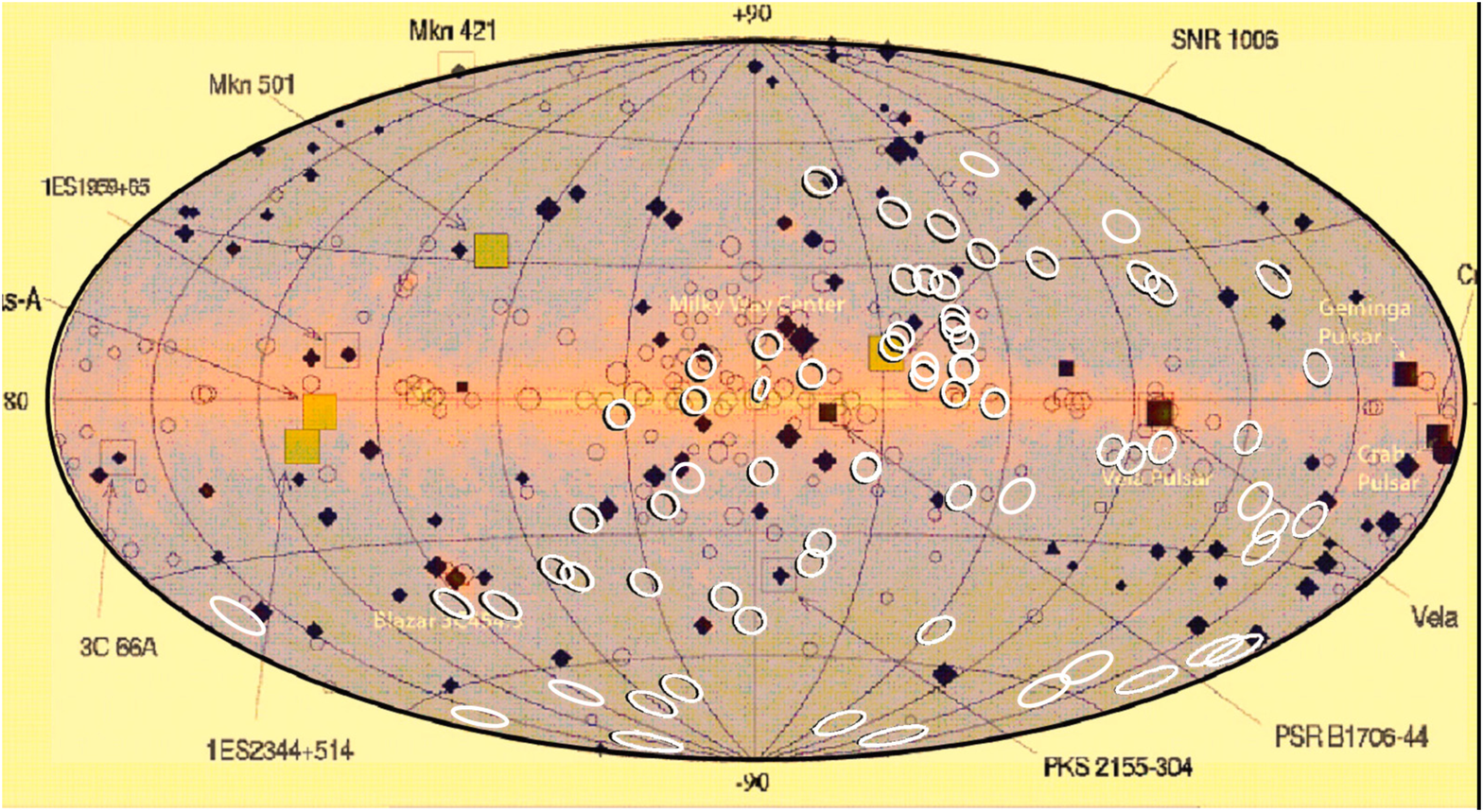}
\vspace{-0.3cm}
\end{center}
\caption{ The old gamma EGRET map with remarkable sources name. In the first image the UHECR seem to partially correlate with the Egret unidentified sources. A galactic component appear quite evident. In the second figure it is worth to notice a few correlated UHECR events rings along the galactic center, the SN1006 source, but also some surprising potential connection of a doublet with far extragalactic AGN as $PKS 2155-304$ and an unexpected far $3C454.3$ whose distance is almost half the Universe size from us. On December 2009 this source was three times brighter than brightest source, Vela. Such a rare connection, well above GZK distance by more than ten times GZK distance recall the revolutionary (but un-fashion) courier role of  UHE neutrino at ten ZeV energy: they may hit onto relic neutrinos on the way (within GZK radius) and lead to Z-shower whose nucleons may be the  observed signal, see \protect\cite{Fargion1997},\cite{Weiler1997};\cite{Yoshida1998}.
}
\label{fig4}
\end{figure}

\begin{figure}[h!]  
\begin{center}
\epsfysize=5cm \hspace{5.0cm} \epsfbox{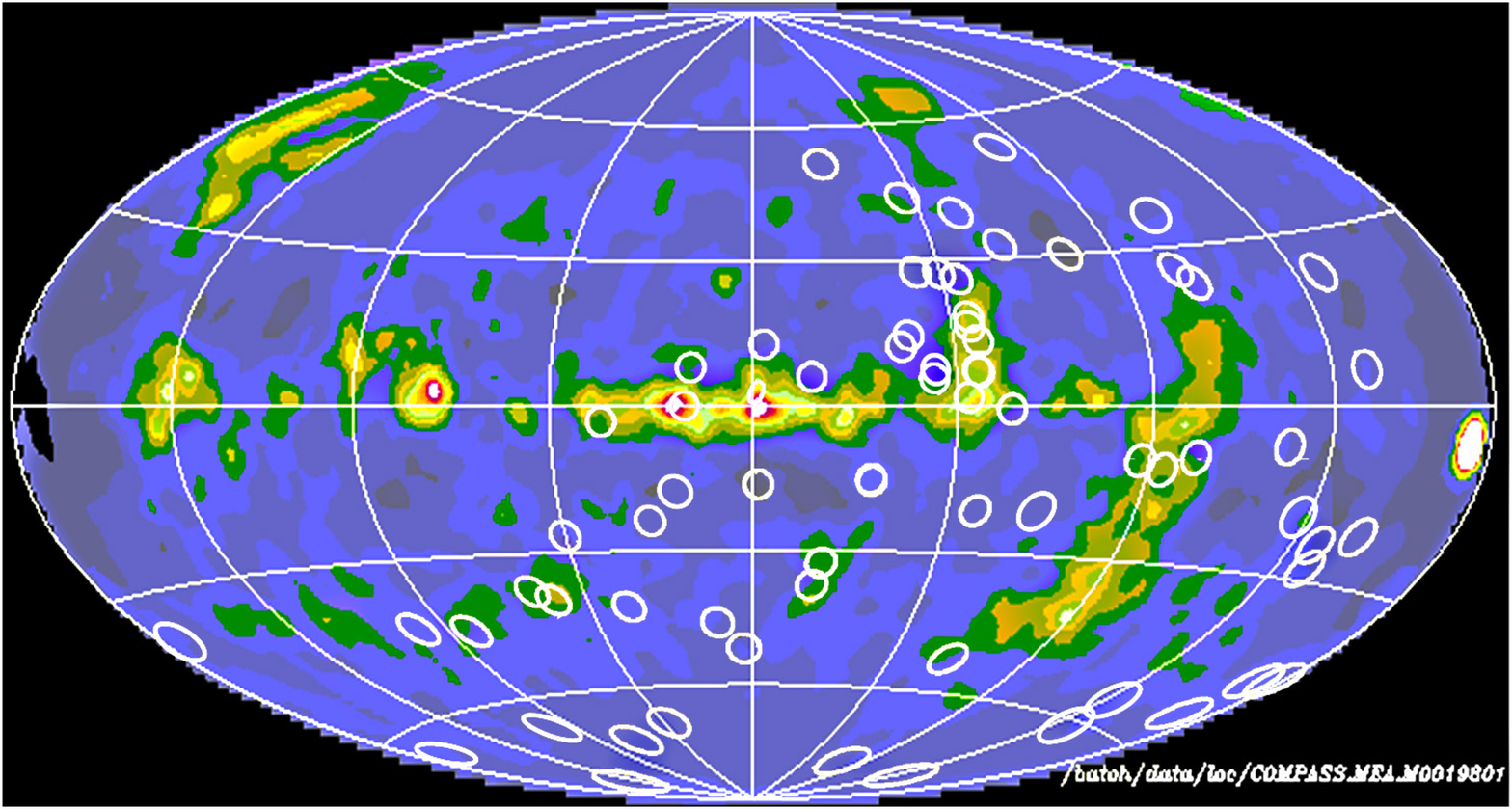}
\epsfysize=5cm \hspace{5.0cm} \epsfbox{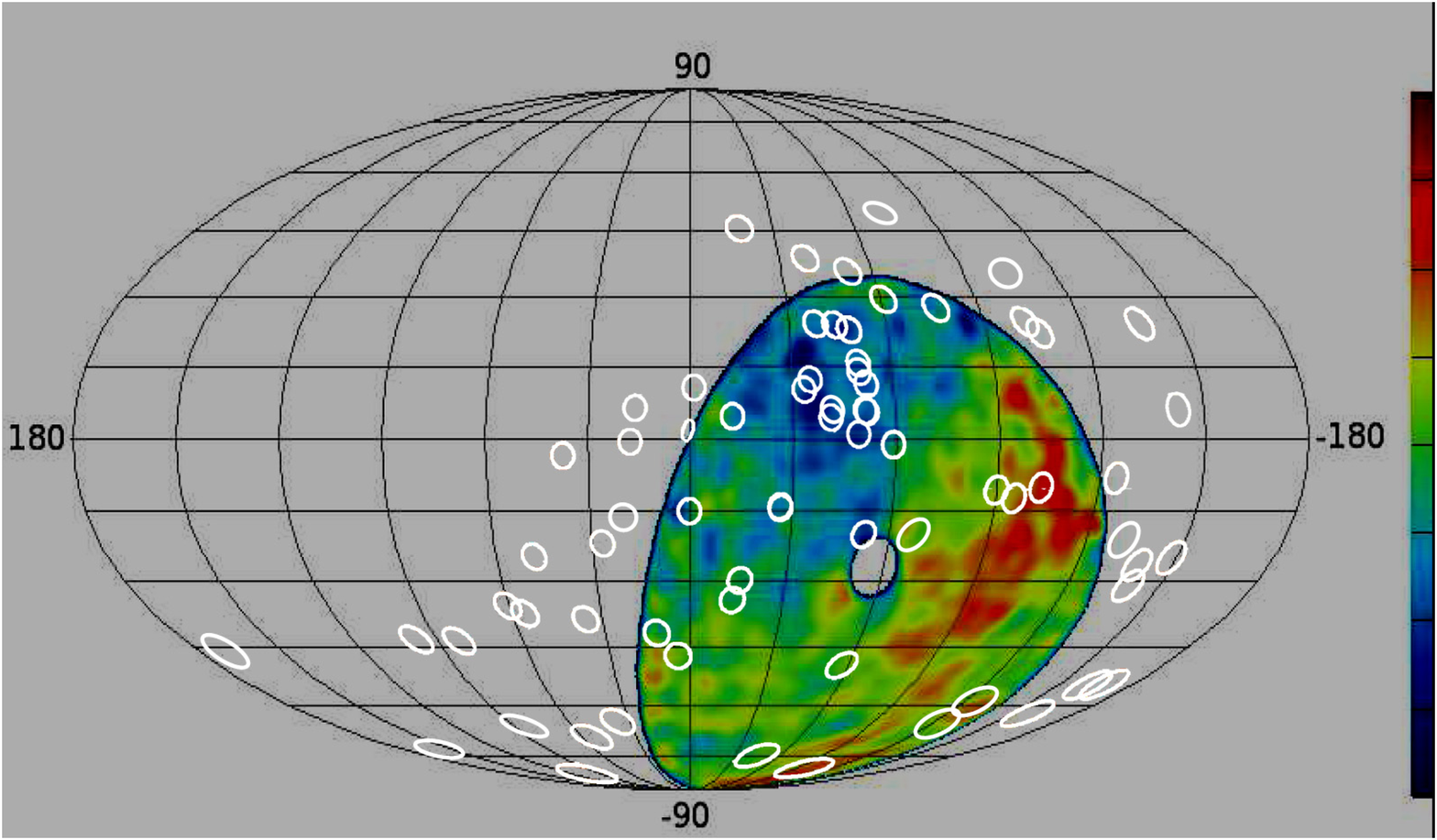}
\epsfysize=5cm \hspace{5.0cm} \epsfbox{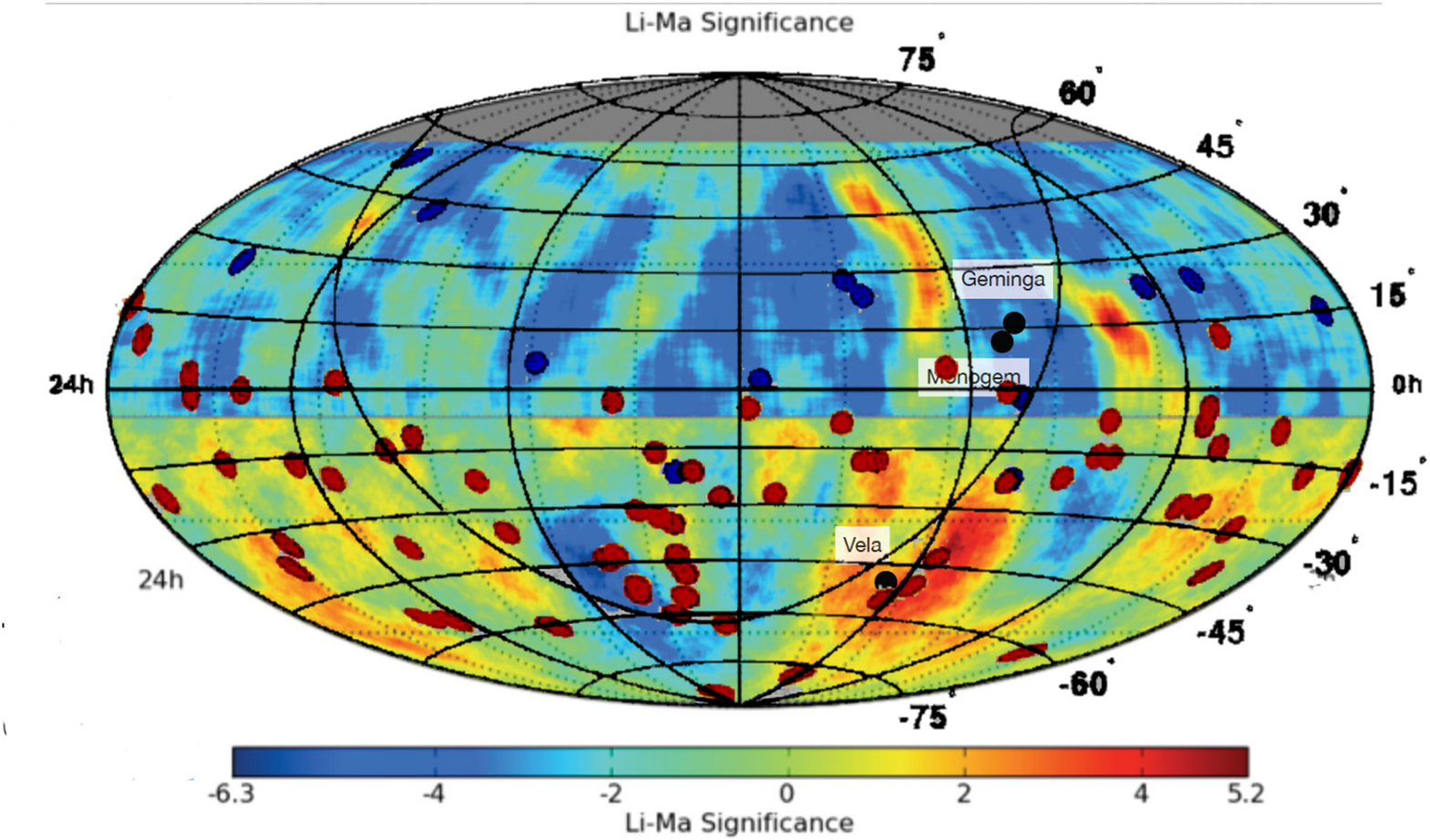}
\vspace{-0.3cm}
\end{center}
\caption{ The old gamma Comptel map (1-3 MeV energy band)  with remarkable correlation area with the 69 UHECR. This few MeV radiation is mostly of galactic sources: Two unique sources that appear in the COMPTEL energy range are isotopes of titanium (Ti-44) and aluminum (Al-26), which are both produced in supernova explosions; decaying aluminum can point to ancient supernova remnants. Other objects in COMPTEL's range include active galactic nuclei (AGN), which are thought to host massive black holes of one million to one billion times the mass of the Sun, mostly (excluding CenA) at distances much larger than GZK. The clustering around CenA and the triplet around Vela, are correlated both in Comptel map as well as in next second map over the new 12 TeV muon anisotropy discovered by IceCube \cite{Desiati}.This hint of evidence is shown both in galactic coordinated (in second figures) as well as in the third one in celestial map. Both Argo and Icecube CR anisotropy are shown at once with the 69 Auger events (orange and red) as well as the 13 Hires (blue) UHECR events.
}
\label{fig5}
\end{figure}

\begin{figure}[!ht]  
\epsfysize=4.2cm \hspace{0.1cm} \epsfbox{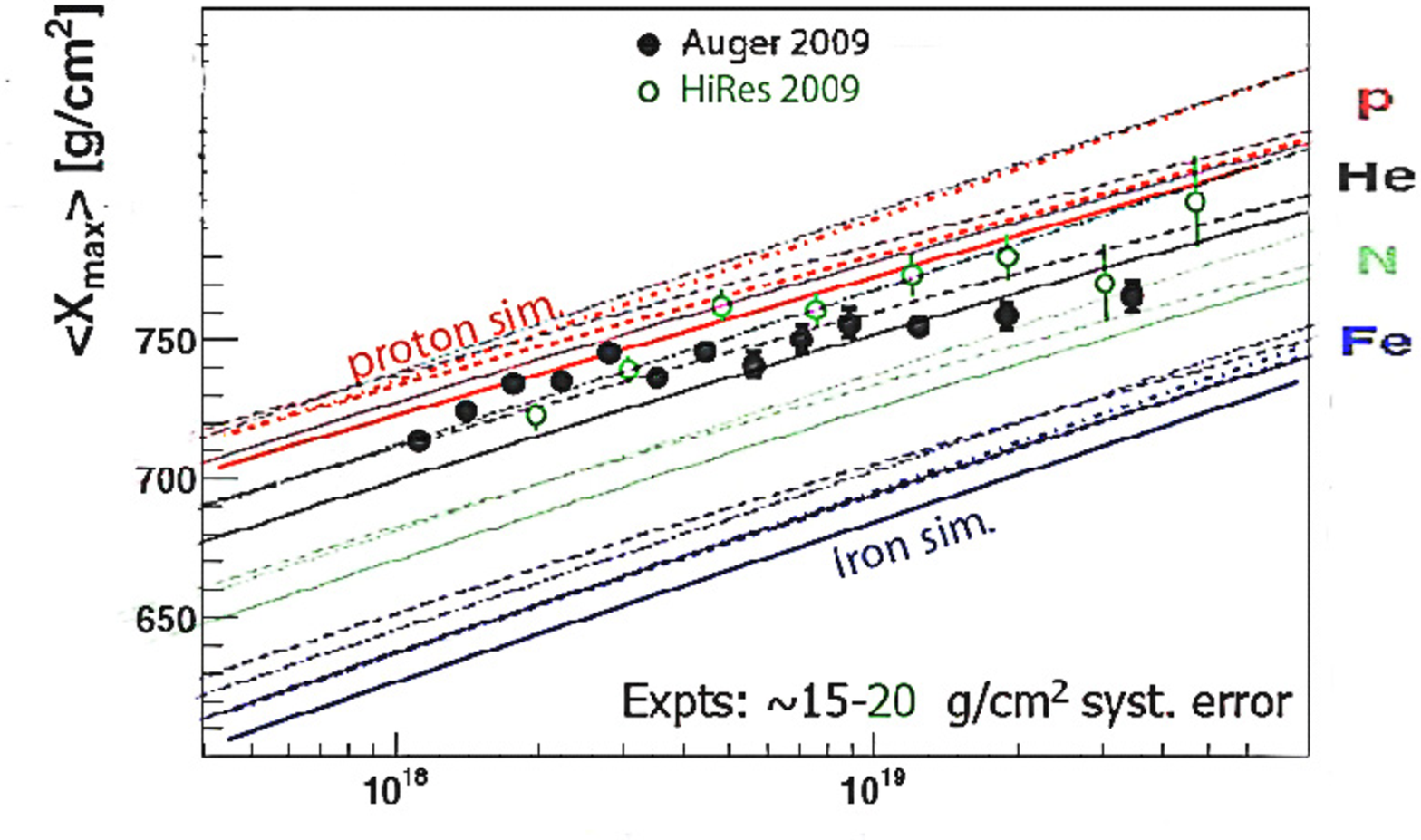}
\epsfysize=4.2cm \hspace{0.1cm} \epsfbox{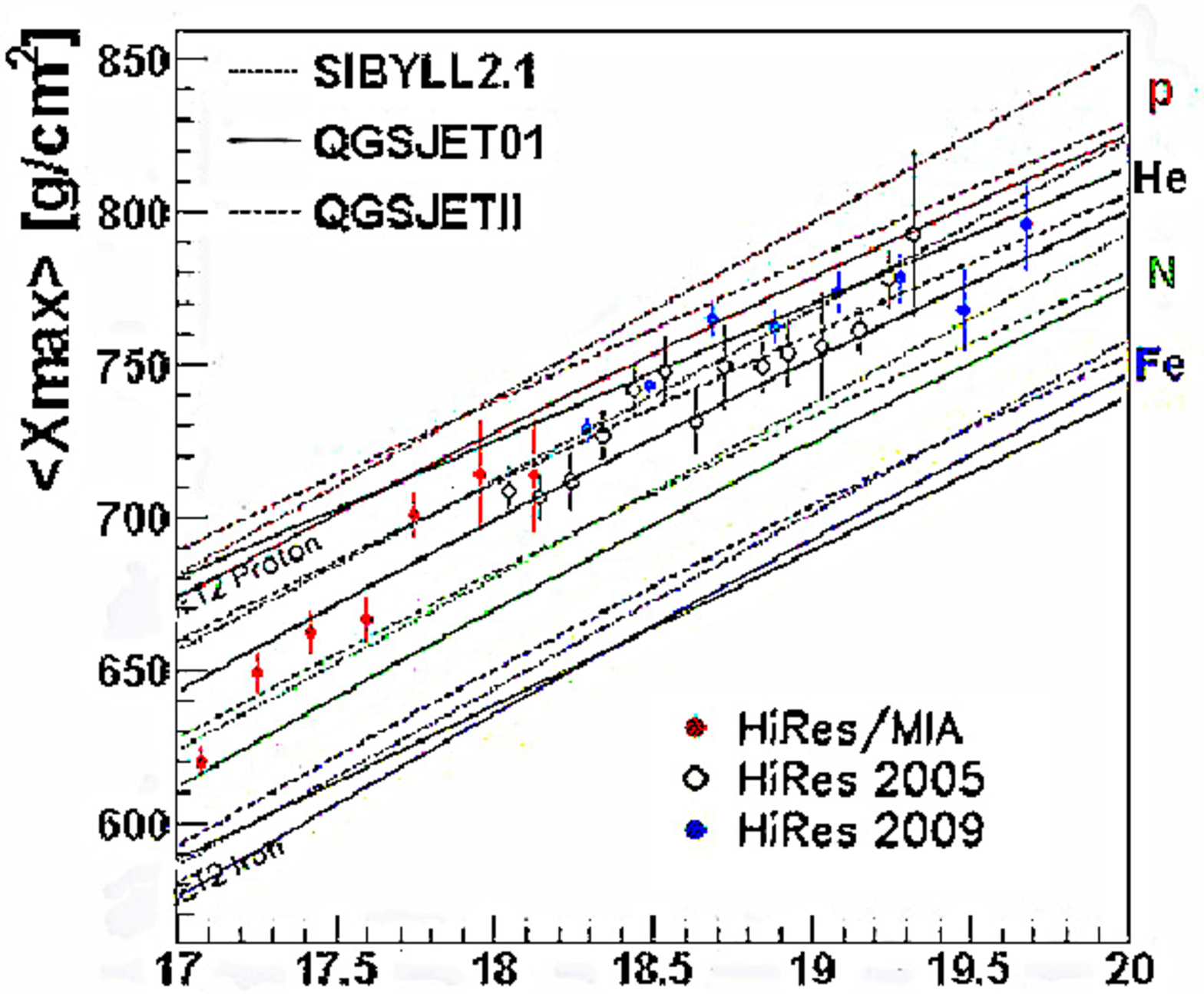}
\vspace{-0.3cm}
\caption{Left: The most recent  UHECR composition map derived by AUGER and HIRES with expected model curves. The He possibility is a better solution also respect an iron-proton mix, because of the narrow error band in AUGER records. Right: The most recent  UHECR composition map derived by  HIRES and Hires-MIA in last decade with expected model curves. The He possibility is a viable solution.
}
\label{fig6}
\end{figure}

\begin{figure}[!ht]  
\epsfysize=4.4cm \hspace{0.1cm} \epsfbox{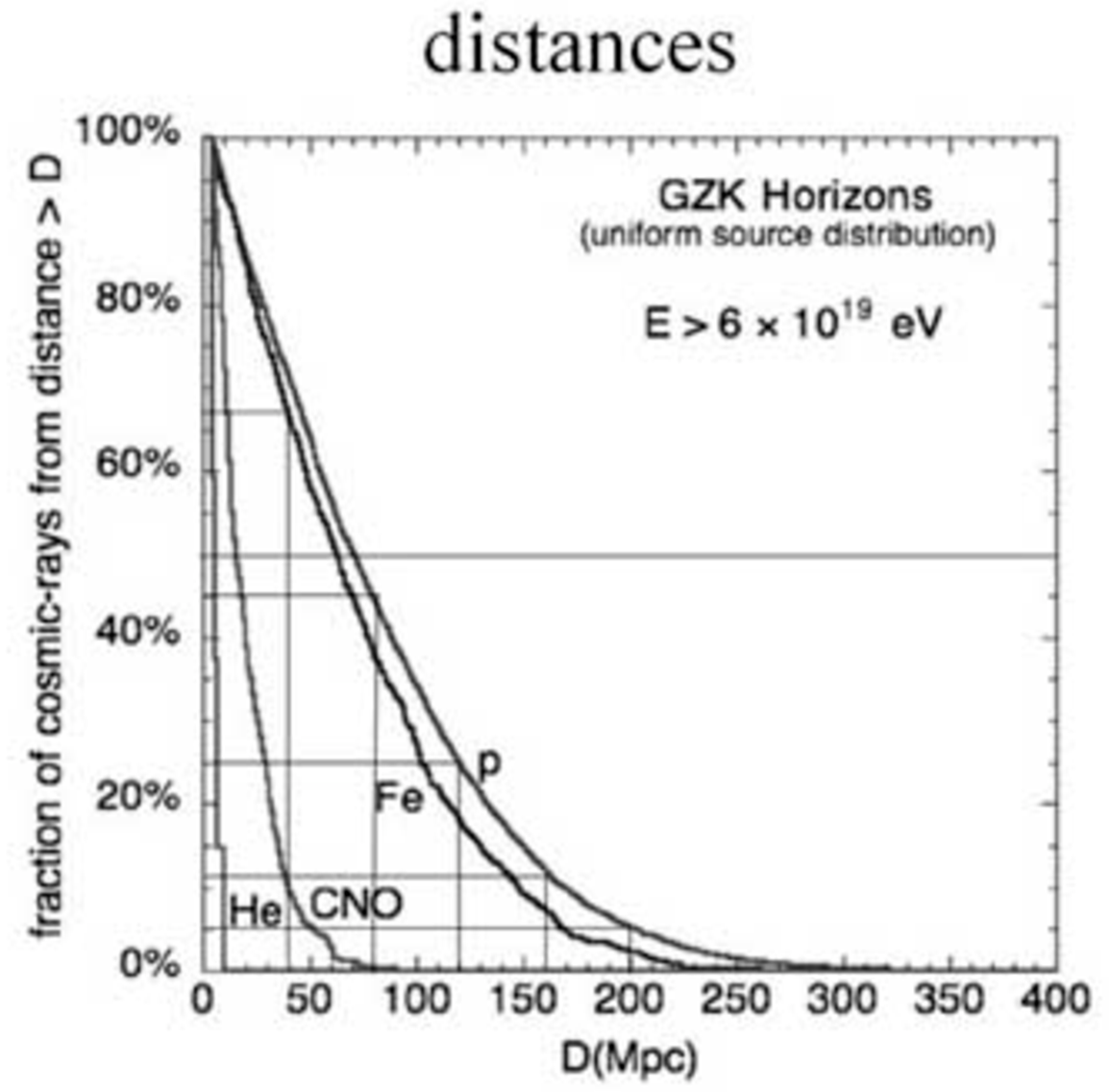}
\epsfysize=4.4cm \hspace{0.1cm} \epsfbox{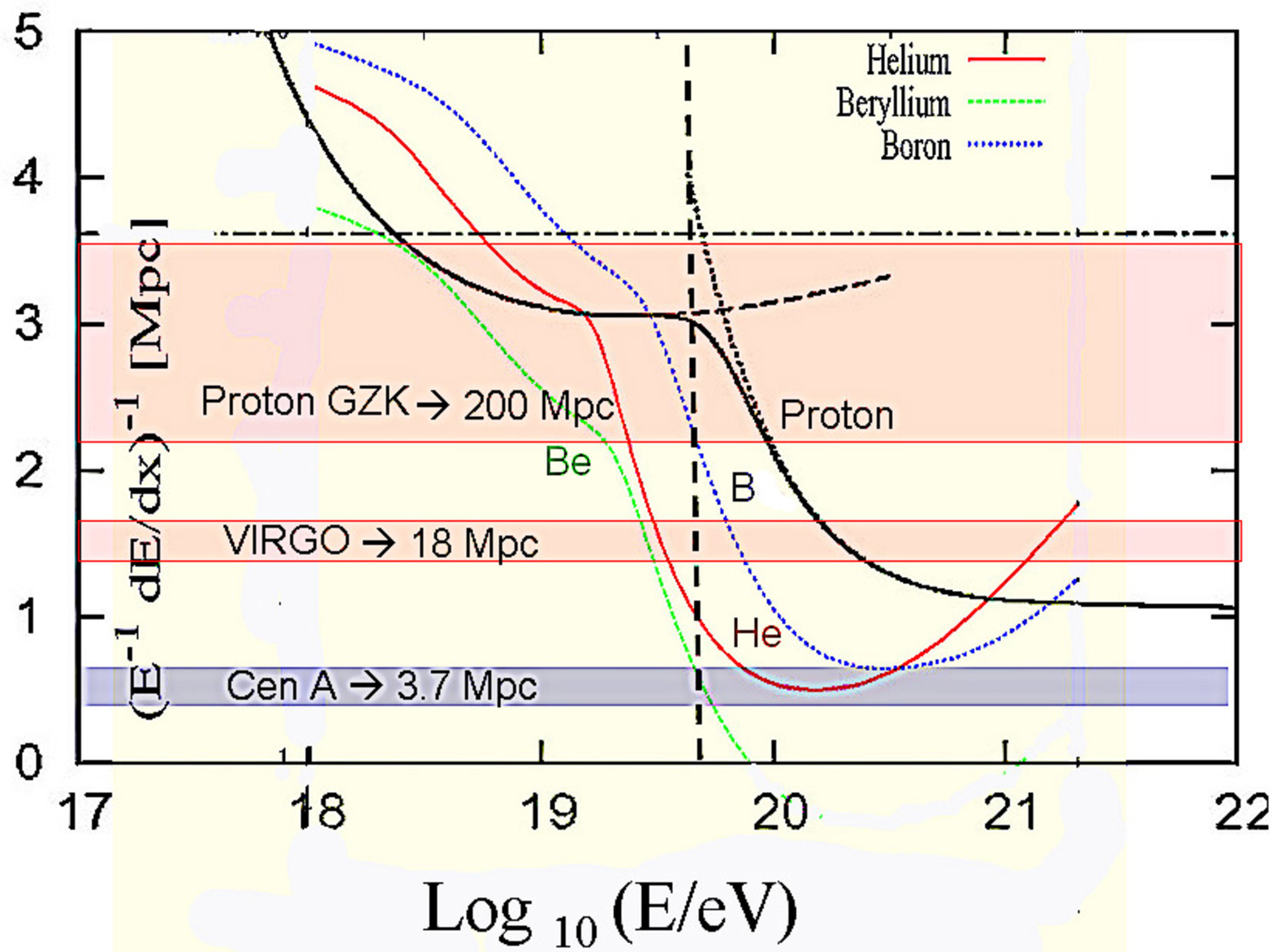}
\vspace{-0.3cm}
\caption{ Left: the suppression fraction of UHECR proton with distance \cite{Auger-Nov09} : this distance dependence explain why UHECR $He^{4}$ are unable to arrive Virgo and why different Infrared maps above are more or less opaque to GZK. The extreme far AGN at far red-shift ($z\geq 0.04$)  are exponentially suppressed by GZK cut-off. This explain the merit of the UHECR originated by UHE neutrino scattering at Z-resonance \protect\cite{Fargion1997},\cite{Weiler1997};\cite{Yoshida1998} to explain eventual correlation with far (much than GZK cut)  AGN.
On the right side the interaction length for proton, Helium and lightest nuclei, showing the main extreme distances. This curve explain why , for Helium like  nuclei, Cen A maybe still observed while Virgo is already isolated from us. }
\label{fig7}
\end{figure}

\begin{figure}[!ht]  
\begin{center}
\epsfysize=5cm \hspace{0.1cm} \epsfbox{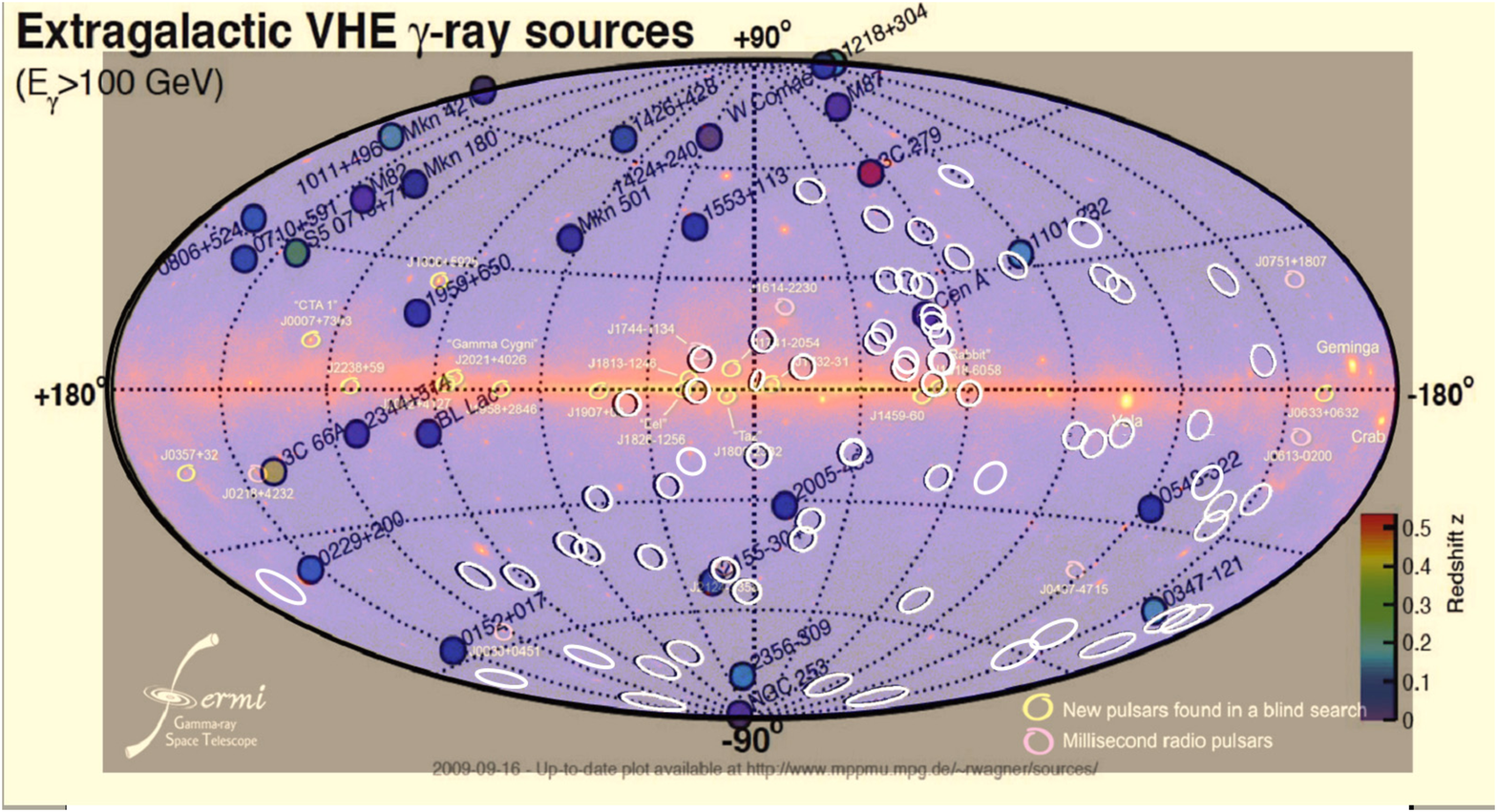}
\epsfysize=5cm \hspace{0.1cm} \epsfbox{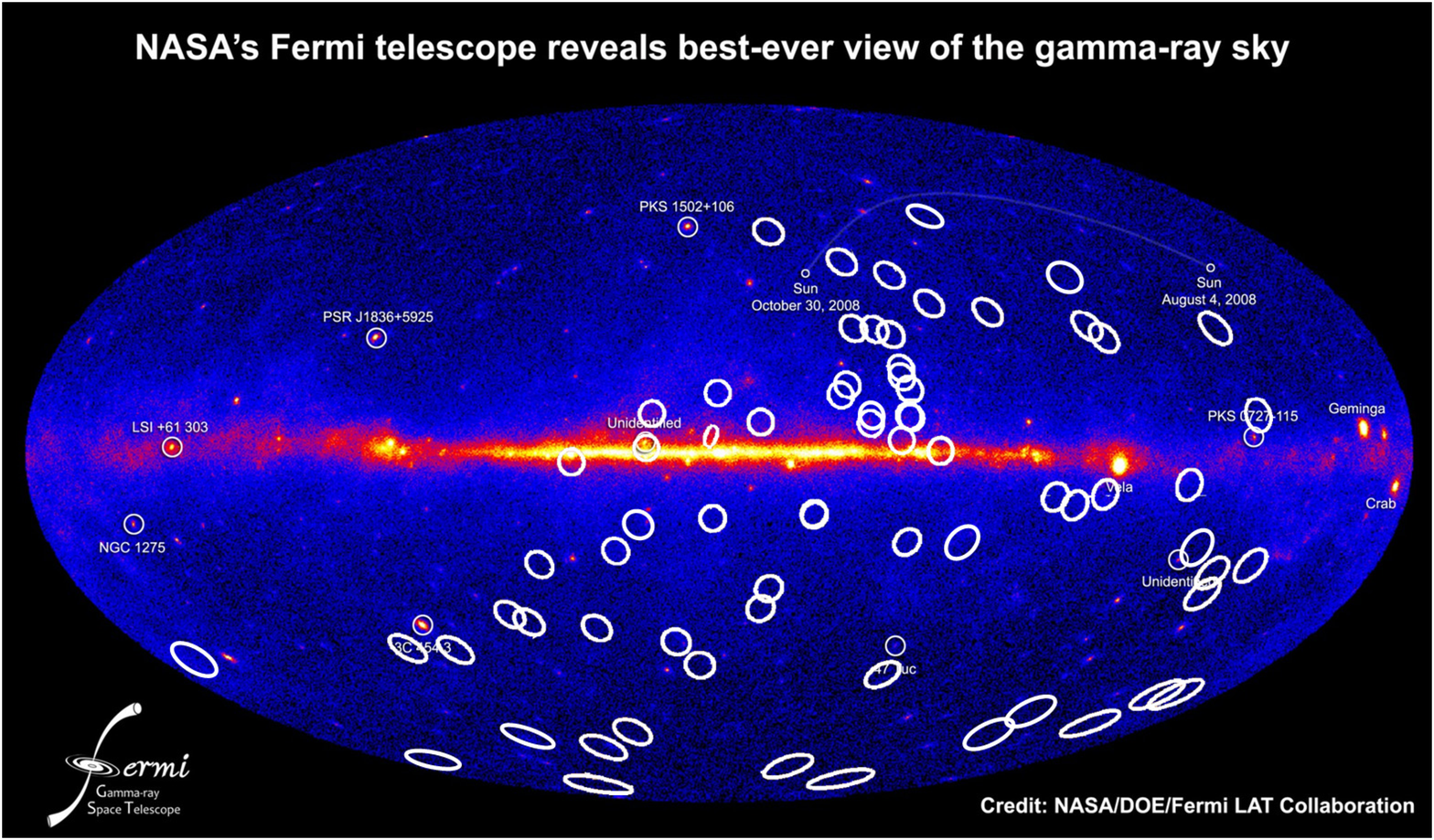}
\vspace{-0.3cm}
\caption{Left: The Hardest Gamma sources versus AUGER maps. Under the image is shown the Fermi gamma sky in low background. In this sky dot map, (whose blue-red colors mark the nearby-far red-shift), one observe the remarkable Cen A correlation with the main UHECR string events. However one is tempted to correlate also the AGN $1101-282$, AGN $0347-121$,AGN $2005-489$,AGN $2155-304$. These sources are (by redshift colors) well above the GZK cut off. The only viable possibility is an extreme UHE neutrino at ZeV energies scattering on relic ones within a GZK volumes \protect\cite{Fargion1997},\cite{Weiler1997};\cite{Yoshida1998}. The very relevant UHECR connection is with blazar $3C454.3$ whose exceptional flare has been discovered just last month. Its distance is half the way the Universe (above two Gpc) requiring (if the correlation is true) a an extreme UHE neutrino at ZeV energies scattering on relic ones. See next figure. Right:The very recent Flare from far AGN blazar $3C454.3$, at half the Universe distance in Fermi sky and with UHECR. Its relevance is not just related to the huge output of the source and to the doublet AUGER event connected by this map: but also to additional signals to be discussed in forthcoming article. The Z-resonant (or Z-Burst) model explains this otherwise mysterious connection. The UHE neutrino primary energy need to be nearly $10-30$ ZeV and the relic neutrino mass might range in the $0.4-0.133$ eV. The whole conversion efficiency might range from a minimal $10^{-4}$ \cite{Fargion1997} for no relic neutrino clustering up to $4 \cdot 10^{-3}$ a (forty) times density contrast in Local Group halo. Even within the minimal conversion efficiency, observed  gamma flaring $3C454.3$ blazar  is consistent with the extreme UHECR flux assuming a primary Fermi flat spectra (of the blazar) extending up to ZeV energy. See \cite{Fargion1997}.
}
\end{center}
\label{fig8}
\end{figure}
\section{Conclusions}
 The history of Cosmic Rays and last UHECR discoveries (and disclaims) are exciting and surprising. The list of models that rose and fall  just last decade is confusing but also promising of new horizons and revolutions. We all hope with courageous experiment fighters on the ground, as Fly's Eye, AGASA, Hires and AUGER ones, are no disappointed and that a new UHECR  astronomy is born, possibly as expected within a GZK Universe \cite{Auger-Nov09}. But Nature is sometime hides its final picture inside inner boxes. The very surprising correlation with Cen A, the absence of Virgo, the hint of correlation with  Vela and galactic center might be solved by a lightest nuclei, mainly He, as a  courier, leading to a very narrow (few Mpc) sky for UHECR. However the very  exceptional  blazar $3C454.3$ flare on $2nd December$ $2009$, a month ago, and the few AGN connection of UHECR far from a GZK volume may force us, surprisingly, to reconsider an exceptional model: Z-Shower one. Possibly connecting lowest neutrino   particle ($\simeq 0.15$ eV) mass with highest UHE ($\simeq 30$ZeV)neutrino energies \cite{Fargion1997}. Even for the minimal UHE $\nu$-Z-UHECR conversion as low as $10^{-4}$ (see table$1$,last reference in\cite{Fargion1997}), \emph{for a not clustered relic neutrino halo as diluted as cosmic ones}  the present gamma $3C454.3$  output (above $3\cdot 10^{48}$ $erg s^{-1}$) is  comparable with UHECR (two events in $4 years$ in AUGER), assuming  a flat Fermi spectra (for neutrinos) extended up to UHECR ZeVs edges. These results are somehow surprising and revolutionary. We might be warned for unexpected very local and a very wide  Universe sources sending UHECR traces in different ways. Testing finally the most evanescent \emph{hot} neutrino relic background. The consequences may be soon detectable in different way by Tau air-showers in AUGER,TA, Heat and also in unexpected horizontal shower in deep valley by ARGO \cite{FarTau},\cite{Auger08}. Or  in widest atmosphere layer on Earth, Jove and Saturn, see 2007 papers in\cite{Fargion1997}. Additional surprises, to be discussed in detail elsewhere, are  waiting beyond the corner. A soon answer maybe already written into present  clustering (as Deuterium fragments) at half UHECR edge energy around or along main UHECR group seed. Better understanding will rise by new data. The next release of UHECR update events, possibly above ten EeV, and their  maps may solve the puzzles. As well the eventual revolutionary discover of a tau airshower nearby (PeVs) or far (EeV) from fluorescence Auger and TA telescopes.

\end{document}